\documentclass[preprint,12pt]{elsarticle}
\usepackage{amssymb}
\usepackage{ifthen}
\usepackage{graphicx}
\usepackage{amssymb}
\usepackage{tabularx}
\usepackage[table]{xcolor} 
\usepackage{hyperref}

\usepackage{multirow}

\newboolean{showcomments}
\setboolean{showcomments}{false}	
\ifthenelse{\boolean{showcomments}}
  {\newcommand{\nb}[2]{
  \fbox{\bfseries\sffamily\scriptsize\colorbox{yellow}{#1}}
    {\sf\small$\blacktriangleright$\textit{#2}$\blacktriangleleft$}
   }
  }
  {\newcommand{\nb}[2]{}
   
  }
\newcommand{\COMMENT}[1]{}
\newcommand\TODO[1]{\nb{TODO}{#1}}

\newcommand\FED[1]{\nb{Fed}{#1}}

\newcommand{\absdiv}[1]{%
\par\addvspace{.1\baselineskip}% adjust to suit
\noindent\textbf{#1}\quad\ignorespaces
}

\journal{ArXiv}

\begin{document}

\begin{frontmatter}


%%\author{Federico Tomassetti\corref{cor1}\fnref{polito}}
%%\author{Marco Torchiano\fnref{polito}}
%%\cortext[cor1]{}
%%\address{Dip.Automatica e Informatica\\Polite\fnref{label3}}

%\title{How MDD can be deployed to an IT ecosystem \\the CSI-Piemonte Case }
%\title{How MDD induces re-shapes an IT ecosystem \\the CSI-Piemonte Case }
%\title{Evolution of a regional IT ecosystem: \\the CSI-Piemonte Case }
%\title{On the IT Ecosystem Reshape Induced by MDD Introduction: \\the CSI-Piemonte Case }
%\title{The reshape of a software ecosystem induced by MDD: \\the CSI-Piemonte Case }
%\title{MDD Induced Reshape of a Software Ecosystem: \\the CSI-Piemonte Case }
%\title{How a Paradigm Shift Reshapes a Software Ecosystem: \\the CSI-Piemonte Case}
\title{A Software Ecosystem Reshaped by a Paradigm Shift: \\the CSI-Piemonte Case}

\author[label1]{Federico Tomassetti}
%\ead{federico.tomassetti@polito.it}
\author[label1]{Marco Torchiano}
\ead{marco.torchiano@polito.it}
\author[label2]{Mauro Antonaci}
\author[label2]{Paolo Arvati}
\author[label1]{Maurizio Morisio}
%\ead{maurizio.morisio@polito.it}

\address[label1]{Politecnico di Torino, C.so Duca degli Abruzzi 24, Torino, Italy}
\address[label2]{CSI Piemonte, Corso Unione Sovietica 216, Torino, Italy}


%\address{}

\begin{abstract}
%Context
\absdiv{Context:}
Changes in the software development paradigm, when operated by entities with a pivotal role, have the power to affect a number of groups and entities in their sphere of influence, changing both their working habits and relations.

%Goal
\absdiv{Objective:}
In this paper we present the organizational changes occurred in a software ecosystem as consequence of a technological change. In particular we examine the evolution of an MDD solution and the changing roles of the company promoting it, the public administrations and the sub-contractors.

%Method
\absdiv{Method:}
The paper focuses on a single case study that encompasses the six years long evolution of a Model-driven development solution, starting from its conception until is recent open-source release, across five distinct phases. The history was analyzed jointly by software engineering academics and industrial managers directly involved in the case study. 

%Results
\absdiv{Results:}
A report of the ecosystem evolution from an idiographic perspective is reported.
An analysis of the history allowed an abstraction that led to the identification of several distinct ecosystem evolution motifs.

%Conclusions
\absdiv{Conclusion:}
The motifs represent a set of key process areas for the evolution of a software ecosystem. They are potentially generalizable to other similar ecosystems.
As such, they can be used by researchers to evaluate existing in-progress case studies, and by practitioners as a set of guidelines.

\end{abstract}


\end{frontmatter}

% \linenumbers

\section{Introduction}
\label{sec:introduction}
A software ecosystem can be defined as: ``a set of actors functioning as a unit and interacting with a shared market for software and services together with the relationships among them'' \cite{Jansen2009}. We believe that the term market should be taken in its most general sense: the set of relationship existing between (type of) actors representing any form of exchange. Typically, in a software ecosystem, actors exchange software artefacts, services, and of course money.  As for a natural ecosystem, also in software ecosystems variations in the behavior of one actor (species) cause reactions from other actors and alterations of the overall environment. In such a kind of complex systems both changes originating from the participants or perturbations coming from external factors (e.g., the economic conjuncture) trigger chain reactions by many of the actors which take part in the ecosystem; the result is either the creation of new relations, modification of the existing ones, or destruction of some of them. Therefore the shape and behavior of an ecosystem as a whole is extremely difficult to predict and govern.

\subsection{Context}

This paper focuses on an ecosystem centered around a large publicly owned organization,
CSI-Piemonte (Consortium for Information Systems), considering the relations between departments, with ten of sub-contractors and with the customers (hundreds). CSI-Piemonte (CSI hereinafter) was founded in 1977 with the aim of promoting the modernization of local administrations by using IT-based tools to create information services and systems. It focuses on the development and operation of  Information \& Communication Technology projects for the Piedmont's Public Administrations (PAs), providing services for citizens and businesses. 

CSI is a consortium with over 100 members, most of which are PAs: the Piedmont Region, several Provinces, and many municipalities. Other members are universities, hospitals, and local health agencies. Many of the members of CSI are also among its customers. Services are developed for many of the PAs of Piedmont which counts over 4.5 millions of inhabitants distributed across over 1200 municipalities (this high number is due to the orography of the region). 

The goal of this paper is to document the evolution of the CSI-centered ecosystem over a span of 5+ years, during the introduction of a new development technology: Model-Driven Development (MDD) \cite{Mellor2003}. The introduction of MDD started in 2008 and is still in progress. The new technology induced several changes in the ecosystem, concerning both the role of the actors and their interactions. In parallel and tightly interlocked with the evolution of the ecosystem we will follow the evolution of the MDD supporting toolset.

\subsection{Motivation}

The presence of a central catalyst and a shared technology, bringing specific benefits to the different participants are fundamental to create a cohesive ecosystem, motivating everyone to favor the success of the technology and, as consequence, benefiting the whole ecosystem. %Committment and support have to be gained first inside the organization boundaries, and then expanded to the rest of the ecosystem.

While the entity at the center of the ecosystem -- i.e. CSI -- was able to build this cohesive ecosystem and earn the support of the participants, still, a huge effort had to be spent to steer the ecosystem and operate a mindset change, winning the inertial resistances. CSI had to initially spend a huge effort not only developing the tools but also investing in complementary aspects (IDE integration, documentation, support, lobbying). However the success in the transition was made sustainable in the long run by the progressive involvement of other actors that helped in a increasingly more active way as progressing in the our story.

Creating the fertile pre-conditions and the determination of the steering organization are however not enough for the survival of the ecosystem. They are complex systems where entities with possibly conflicting goals and a number of inter-relations co-exist. In this kind of environment  technology could play the role of the enabler for a mindset change but many other aspects are crucial and have to be properly considered: among them we wish to underline necessary competencies, organizational aspects, economic aspects. In such complex systems, where so different aspects have to be considered, it is hard to forecast the effects of changes and the long-term results of actions. This could lead to unanticipated benefits (like the spreading of MDD competencies in a local area) but to problems as well. We therefore think that steering actors or simple participants in similar ecosystems could benefit from a few guidelines, emerging from successful cases of paradigms transitions operated in software ecosystems.

\subsection{Organization of the paper}

The paper is organized as follows: section \ref{sec:method} introduces the method adopted to conduct the study, section \ref{sec:story} describes the evolution from an historical perspective, section \ref{sec:motifs} analyzes and distills different motifs that characterize the successful transformation of the ecosystem. Later we discuss the main aspects of the transformation of the ecosystem (section \ref{sec:discussion}), then we present the related works (section \ref{sec:relwork}) and finally we draw our conclusions (section \ref{sec:conclusions}).

%\FED{Todo}
%In Sect. \ref{sec:story} we document the steps through which Model-Driven development (MDD) was adopted in CSI. Results are shown in Sect. \ref{sec:results} while discussion on organizational aspects is contained in Sect. \ref{sec:motifs}. Finally we draw our conclusions \ref{sec:conclusions}.

\section{Method}
\label{sec:method}

The collection of information was carried on over a period of two years starting in April 2011. At that time a collaboration between the Software Engineering group at Politecnico di Torino and CSI Piemonte started, which focused on the development of a model versioning infrastructure to be used for the MDD tool suite.

During the collaboration the researchers became aware of the articulate history related to the conception, development, introduction, and deployment of the MDD solution and decided to undertake an additional hermeneutical research effort focused on the evolution of the ecosystem centered around the MDD tool suite. The team -- the authors of the present paper -- is composed of a group of academic researchers and a group of industrial members.

The research method adopted in this work is essentially of interpretive nature~\cite{KleinMyers1999}. In particular the investigation is based on a single case study that lasted almost six years, which encompasses several hundreds individual software development projects.

The collection of materials occurred in several different occasions.
\begin{itemize} 
\item An initial series of meetings approximately taking place with bi-weekly frequency, they where originally intended to understand the architecture of MDD tools for the purpose of collecting the requirements for the model versioning infrastructure. Those meetings provided an initial overview of the ecosystem and its historical evolution.

\item A workshop was organized in July 2011 for the announcement of the release of MDD-tools as open source software, the researchers participated in this workshop where accounts of experience with MDD tools by third party developers were presented. The feedback from external subcontractors allowed us to confirm the information collected from within CSI.

\item Two meetings were organized to focus on the historic-technical perspective of the ecosystem, the meeting were conducted by the researchers in the form of unstructured interviews where the industrial participants were asked in general about the ecosystem evolution and specifically about the socio-technical aspects that characterized it.
 
\item Eventually a working document was produced to summarize what emerged during the focused meetings and served as the reference for discussion and clarification, which took place mainly via email and telephone.
 
\item After the first version of the paper was available, the researchers conducted two semi-structured interviews with developers who actually used the MDD tools in order to confirm or refute the interpretation.
\end{itemize}

The goal of our investigation is to describe a complex and large local ecosystem and document the main patterns that emerge during its historical evolution. Though our approach is similar to a case study \cite{runeson2009guidelines}, the breadth, temporal duration, geographical extension, and moreover the interpretive nature make it depart from the usual understanding of that kind of study.

%Concerning the detailed research method of work, we adopted an iterative and incremental approach in interpreting the collected materials and to present them in the form of a historical narration of the ecosystem evolution. The 

The type of data collected in the previous events range from informal notes taken on paper, to detailed notes taken e.g. with text editors, to structured notes taken e.g. with mind-mapping tools, to interview transcripts. Due to the heterogeneity of the materials and the mainly interpretive nature of our work we decided not to use common qualitative methods (e.g. coding techniques)\cite{Seaman99}, which are more suitable for hypothesis confirmation -- i.e. a positivist approach -- and homogeneous materials.

We summarize the main features of our work with reference to the basic principles of interpretive field research proposed by Klein and Myers \cite{KleinMyers1999}:

\begin{description}
\item[Fundamental Principle of the Hermeneutic Circle:] it is assumed that ``movement of understanding is constantly from the whole to the part and back to the whole''~\cite{Gadamer1976b}. Our understanding started with specific technological aspects concerning model lifecycle and extended to the whole ecosystem, the several iterations allowed us to achieve a satisfying comprehension.

\item[Principle of Contextualization:] in reporting the evolution of the ecosystem we strive to provide as much contextual information as possible to show how the observed phenomena emerged.

\item[Principle of Interaction Between the Researchers and the Subjects:] several concepts presented in the paper emerged through the interaction among the researchers and the industrial authors. The very idea of considering the MDD tools, its support team, the development teams using it as an ecosystem emerged during the meetings and was not originally present in the data. 

\item[Principle of Abstraction and Generalization:] the first part of this paper (see section \ref{sec:story}) reports and idiographic interpretation of the CSI ecosystem evolution, while the following part (see section \ref{sec:motifs}) attempts to abstract a few general patterns that may attain a wider, nomothetic validity. We do not claim any statistical representativeness of our generalization, its validity relies ``on the plausibility and cogency of the logical reasoning used in describing the results from the cases'' \cite{Walsham1993} and in the substantial agreement with findings from other works available in the literature.

\item[Principle of Dialogical Reasoning:] given our role of authors, we may declare the na\"ive expectation of the adoption of MDD tools to spread though the ecosystem solely due to its pure technical excellence. We actually confronted such preconception in two ways: when data actually supported it we looked for some additional objective measure, on the contrary we reported other factor affecting the diffusion of MDD as they emerged from the analysis of data. 

\item[Principle of Multiple Interpretations:] the main source of information for this investigation relies on the two industrial coauthors who both belong to the MDD support group. Additional viewpoints come from the third party developers presentations during the workshop and from two interviews with individual developers belonging to the group of MDD tools users.

\item[Principle of Suspicion:] the researchers' group often discussed about the information provided by the industrial side and always concluded that the trust relationship was well deserved.

\end{description}

The assessment of the validity of the results is an important methodological part in any experimental and more in general positivist research. We intend to stress that our effort is essentially of interpretive nature, therefore the generalization effort cannot be evaluate in terms of external validity. The motifs reported in section represent an attempt to abstract the key features of the ecosystem evolution, though it is not possible to claim their general applicability to other contexts. They can be compared to existing results in the literature and to the authors' experience, and it is possible to argument -- though not prove -- their potential validity in similar settings.

\section{History}
\label{sec:story}

\begin{figure*}[tb]
    \centering
    \includegraphics[width=1\textwidth]{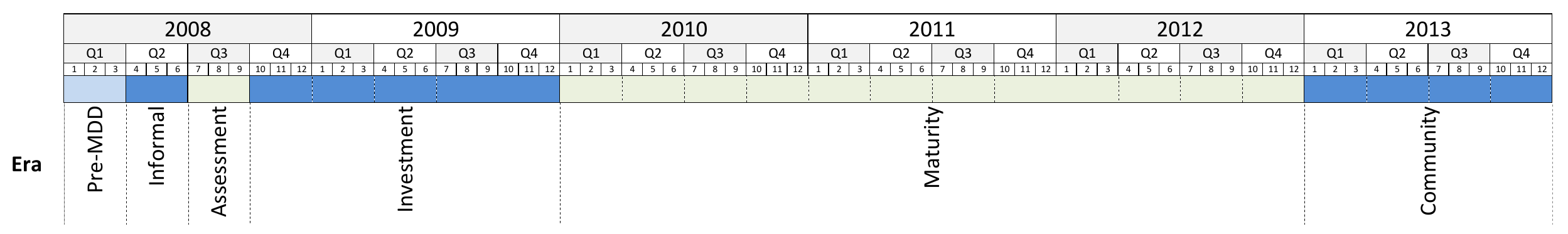}
    \caption{Overall timeline}
    \label{fig:timeline}
\end{figure*}

The software ecosystem revolving around CSI started its evolution, due to the introduction of the MDD approach, back in 2008; we document its evolution over a five years period until 2013, time of this writing. We start by presenting the environment before the introduction of MDD and then document the evolution through five main eras. Figure \ref{fig:timeline} shows the overall chronological layout of the eras:
\begin{itemize}
\item {\bf Informal:} initial MDD tools are developed and used in an ad-hoc manner;
\item {\bf Assessment:} commitment to evaluate a possible platform for enterprise-wide adoption;
\item {\bf Investment:} development of MDD-Tools platform and diffusion within enterprise boundaries;
\item {\bf Maturity:} involvement of contractors and enhanced support;
\item {\bf Community:} the community take responsibility for the MDD-Tools.
\end{itemize}

%We present in the following the detailed description of the history focusing on how the roles of the different actors changed through the consecutive phases. %\FED{Questo non e piu vero per results, giusto?} Results and discussion are part of the next sections.

In order to represent the ecosystem throughout its evolution we use a set of diagrams -- one per era -- that represent the members of the ecosystem as ovals. The artifacts and services exchanged by the members are reported as arrows from the producer to the consumer.
We depict sub-systems -- i.e. groups of members that play a specific role -- using dashed ovals. Sub-systems can be constituted by single working groups (e.g., the software engineering group), by categories of personnel (e.g., business analysts) or categories of organizations (e.g., sub-contractors). For every era we highlight with thick (red) lines the changed or new elements with respect to the previous era.

%\FED{Ho provato ad aggiungere questo} 
%\MTK{da spostare in intro?}
During the different eras it is possible to observe how the new paradigm starts as an autonomous initiative of a small group of developers (\emph{Informal}), then it gains the attention of the CSI management (\emph{Assessment}), it is adopted within the boundaries of the CSI organization (\emph{Investment}). Later the change affects the whole ecosystem (\emph{Maturity}) and finally the responsibility for the solution is moved away from the central organization to be distributed across the ecosystem (\emph{Community}).

\subsection{Before MDD Era}

\hfill {\bf Period:} \emph{until March 2008} \vspace{1em}

\begin{figure*}[tb]
    \centering
    \includegraphics[width=.75\textwidth]{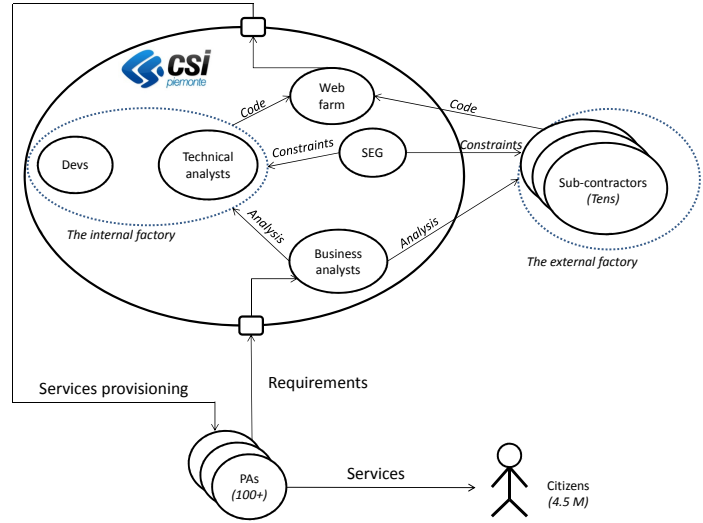}
    \caption{Ecosystem before MDD}
    \label{fig:ecoBeforeMDD}
\end{figure*}

The initial configuration of the ecosystem can be observed in Figure \ref{fig:ecoBeforeMDD}.

In CSI different kinds of applications are developed. Among them a large number are web applications to be plugged into web portals for several different public administrations. These web applications often need to comply with rigid internal rules, which include coding standards, building standards, graphical standards, and rules for other technical aspects such as the mechanisms for authentication. The standards are in place to make the development process more robust and reliable, obtaining predictable costs, and ensuring the products meet given quality standards. Development rules are mainly defined by the Software Engineering Group (SEG).

The presence of rules and conventions is also motivated by the fact that CSI does not offers software development services only, but provides hosting to some of its larger institutional customers too. Thus the company aims to produce applications based on small number  of technological infrastructures, with the goal of reducing the effort for deployment and maintenance of the web farm. The web farm is an infrastructure managed by CSI and used to host a large number of customers' web applications.

The organization strives to have a software factory where the kind of applications typically developed (web applications) can be produced according to a precise and repeatable process. The software factory is partially internal (consisting of technical analysts and developers employed directly by CSI) and partially external (i.e. sub-contractors). %\TODO{E' da chiarire che sono intercambiabili?}
% Marco: direi di no � essenziale per il resto?

Customers describe requirements to business analysts from CSI; who produce an analysis document that is delivered to the \emph{software factory}. The factory builds the application, according to the analysis and adhering to the development rules. %The software factory is constituted by developers employed by CSI and by sub-contractors. 
Once produced, the application is typically passed to the system admins who are responsible for the deployment in the internal web farm.

The reported baseline productivity -- at this stage of the ecosystem evolution -- was 15 function points per person-month, considering the overall project. While, focusing on the bare development activities (within the software factory), the productivity was 30 function points per person-month. 
For the computation of functional size CSI uses IFPUG function points \cite{2012ifpug}. 
%\MTK{in CSI utilizziamo o le metriche IFPUG o gli Early \& Quick FP (la seconda tecnica principalmente nei casi in cui si voglia avere un dimensionamento gia in fase di offerta). Mi sembra di ricordare che negli esempi in questione i conteggi siano stati fatti a posteriori (non a scopo predittivo) }

%\subsection{Phase 0: Informa}
\subsection{Informal Era}
\label{sec:phase0}
%\begin{figure}[tb]
%    \centering
%    \includegraphics[width=0.6\textwidth]{images/UML_Diagrams/Phase0_composed2.png}
%    \caption{Phase 0}
%    \label{fig:phase0}
%\end{figure}

\hfill {\bf Period:} \emph{between April 2008 and June 2008.} \vspace{1em}

%\subsubsection{Events}
\begin{figure*}[tb]
    \centering
    \includegraphics[width=0.75\textwidth]{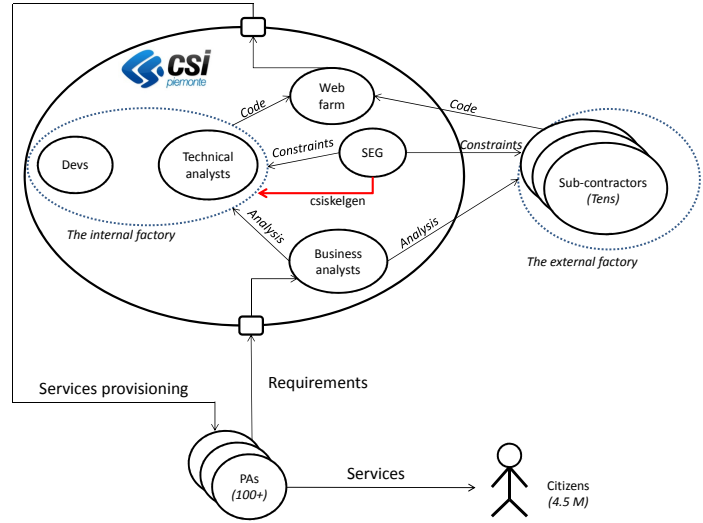}
    \caption{Ecosystem during the Informal era}
    \label{fig:eco0}
\end{figure*}
The SEG developed a tool to generate the skeleton of services. The ultimate goal was to simplify the startup process of projects developing new services and to reduce both the effort and the errors -- frequently due to a {\em copy \& paste} approach. %The tool permitted to reduce the time spent preparing the technological infrastructure, letting more effort be devoted to the application specific features of the service.

The tool was named \emph{csiskelgen} and it was initially intended to be a simple template-based generator. During the development the SEG realized that they could produce a more general and therefore useful tool with a small additional effort. For this reason they decided to make it more flexible adopting a Domain Specific Language (DSL)\footnote{A Domain Specific Language is a language with a limited expressiveness designed to express concisively a certain aspect. Famous DSLs include HTML, CSS and Latex. Recent tools highly reduced the cost of developing this languages, augmenting significantly the number of practicioners realizing them. For more details see \cite{Fowler2010}} for the description of the services to be generated.

%The most common model-driven technologies were evaluated; in particular the SEG focused on EMF and OpenArchitectureWare\footnote{OpenArchitectureWare was later absorbed into the Eclipse Modeling Project}. 
The following two usage modalities were supported:
\begin{description}
\item[basic modality] generate the skeleton of the service and from there develop the application without further using the tool,
\item[advanced modality] define through a model the interface of the service and some aspects like security and transactionality. From this model code could be obtained and application-specific logic could be written within protected regions. Then the model could be later modified and the code re-generated with the protected regions preserved. So the tool could be used along all the duration of the project. Users do not just generate an initial skeleton but they build a model of the service then they progressively refine it during the whole life of the project.
\end{description}

At this stage the company management was largely unaware of the sperimentation with MDD techniques: this was regarded as a technical implementation detail, known only to developers using it. This was possible because project management let the developers pick their own tools and this stage those tools were not regarded as an important asset or a mean to execute a organizational transition.
So the developers working in the company could freely decide whether to use the tools. A large portion of them decide to adopt the "advanced modality".

%\subsubsection{Ecosystem}

This first era of the ecosystem is described in figure \ref{fig:eco0}. The main novelty is represented by the {\em csiskelgen} that is provided by the SEG to the developers in the software factory. 
An important {\em motif} in this era is represented by the adoption of the MDD approach, conducted on a voluntary base (see Sect. \ref{sec:motifIncrementalAdoption}).

No precise data is available about productivity but the developers reported the impression of benefiting from adopting MDD. Actually the diffusion at this stage was primarily driven by an immediate productivity gain perception.

Still some developers did not like to give up the full control on the code but they felt the trade-off was worth while because in return they were shielded from some technical details about more difficult and less creative aspects, e.g. securitization.

%The emerging success of the initial version of the tools made evident the need for integrating the tool within the an IDE and start thinking about a more organic solution.

\subsection{Assessment Era}
\label{sub:phase1}
\hfill {\bf Period:} \emph{between July 2008 and October 2008.} \vspace{1em}

%\MTK{Si potrebbe fondere con la fase precedente, in quanto l'ecosistema non cambia, sono evoluzioni tecniche e gestionali interne al CSI}
%\FED{Forse si potrebbe lasciare (magari tagliando un po') per mostrare i passaggi della diffusione interna (primo passo developer-driven, secondo \MTK{Questa era potrebbe essere fusa con la precedente ma abbiamo scelto di lasciarla a s\'e per mostrare i passaggi della diffusione interna (primo passo developer-driven, secondo management-driven)}
%\begin{figure}[tb]
%    \centering
%    \includegraphics[width=0.6\textwidth]{images/UML_Diagrams/Phase1_composed.png}
%    \caption{Phase 1}
%    \label{fig:phase1}
%\end{figure}

\begin{figure*}[tb]
    \centering
    \includegraphics[width=.75\textwidth]{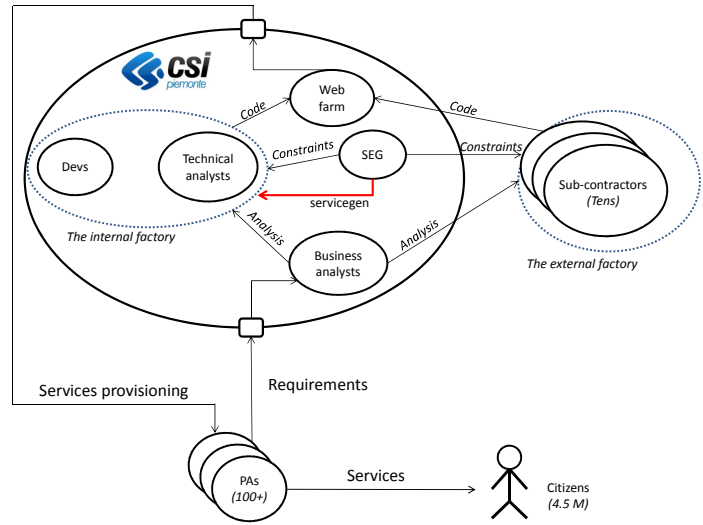}
    \caption{Ecosystem during the Assessment era}
    \label{fig:ecoAssessment}
\end{figure*}

%\subsubsection{Description}

%Figure \ref{fig:phase1} shows the actors -- the same as in the previous phase -- and their involvement during this phase.

In the previous era a basic MDD solution was started autonomously by developers and began to spread exclusively because of word-of-mouth. At this point the CSI management become aware of this solution and decide to conduct an evaluation of MDD technologies. The goal of the evaluation is to drive possible new investments on a more complete toolset.

%MDD platform for the generation of business web applications -- a type of applications frequently realized at CSI --, the evaluation was set to last for a few months. 

In this timeframe the SEG evaluated different software solutions for the rapid development of business web applications. They reached the conclusions that bending those tools to obtain applications as expected by the CSI infrastructure (CSI does not only develop applications abut also host them in its own web farm) would have been very difficult. The company realized the only feasible solution was to craft its own tools. It was clear upfront that the investment to be done would have been relevant. %This motivated the experimentation, to investigate risks and benefit with a limited effort before deciding for full commitment.
%\MTK{Non sequitur: Se l'esito della valutazione � stato contemporaneo allora l'esito negativo non pu\'o essere stata una motivazione}
%\FED{Questa è la motivazione per QUESTA fase, spostato a inizio sezione}
%The technologies finally chosen for the experimentation were EMF and OpenArchectureWare, the same used to develop \emph{csiskelgen}.

The  SEG decided to initially focus on the evolution of \emph{csiskelgen} tool before facing the more challenging task of building a full-fledged web application generation tool chain. The evolution of \emph{csiskelgen} was named \emph{servicegen}. The major enhancements introduced were:
\begin{itemize}
\item the model editor was improved: while before the reflection-based editor for EMF\footnote{EMF stands for Eclipse Modeling Framework (see \url{http://www.eclipse.org/modeling/emf/}. It includes a complete suite of interoperable components to build personalized MDD solutions} models were used a specific one was now developed; %\FED{Mauro,Paolo: confermate?}
\item only the "advanced mode" -- supporting a completed MDD round-trip process\footnote{By the term \emph{MDD round-trip process} we indicate the typical development process which starts by defining models, then generate code from them, permit the manual customization of special region of the generated code and going back to modify again the models, re-generating new code from them without losing the manual customization of the code. This is a circular, iterative process which starts from models and permit to later operate again on models. It is the alternative to the one-shot generation approach in which code is generated from models once and later always edited directly, without further operating on the models.}  -- was kept while the "one-shot" approach was discontinued;
\item the tool was integrated into Eclipse, thus providing a complete platform and uniform user experience;
\item service orchestration modeling was added, covering an aspect traditionally considered hard by developers.
\end{itemize}

%\subsubsection{Results/Ecosystem}
Figure \ref{fig:ecoAssessment} shows the ecosystem during this era. At this level the only noticeable difference is the shift from {\em csiskelgen} to {\em servicegen} but another crucial aspect is that the support for the initiative is growing inside the company boundaries and spread from developers to the management. Moreover something important happened under the hood: enough confidence was gained about the maturity of the enabling technologies, the decision was made in favor of toolsmithing, and the technological platform -- Eclipse EMF -- was selected.
In particular the decision of developing their own tools (see Sect. \ref{sec:motifToolsmithing}) represented a fundamental decision for the future evolution of the ecosystem.

The solution emerging was based on the definitions of a set of meta-models using the Ecore metamodel. Ecore is the solution typically used to describe meta-models when using the EMF platform. It is also the reference implementation of the OMG's EMOF (Essential Meta-Object Facility). The EMF models describing the single applications (and adhering to the designed meta-models) where developed through specific editors created in these phase. From models source code was generated using Xpand\footnote{\url{http://www.eclipse.org/modeling/m2t/?project=xpand}}, a templating system interoperable with EMF models.

In terms of adoption of the MDD approach became more popular, but still limited to services and in particular to orchestrated services.

%\subsection{Phase 2: Investment}
\subsection{Investment Era}
\label{sec:phase2}

\hfill {\bf Period:} \emph{between October 2008 and December 2009.} \vspace{1em}

%\begin{figure}[tb]
%    \centering
%    \includegraphics[width=0.6\textwidth]{images/UML_Diagrams/Phase2_composed.png}
%    \caption{Phase 2}
%    \label{fig:phase2}
%\end{figure}

\begin{figure*}[tb]
    \centering
    \includegraphics[width=.75\textwidth]{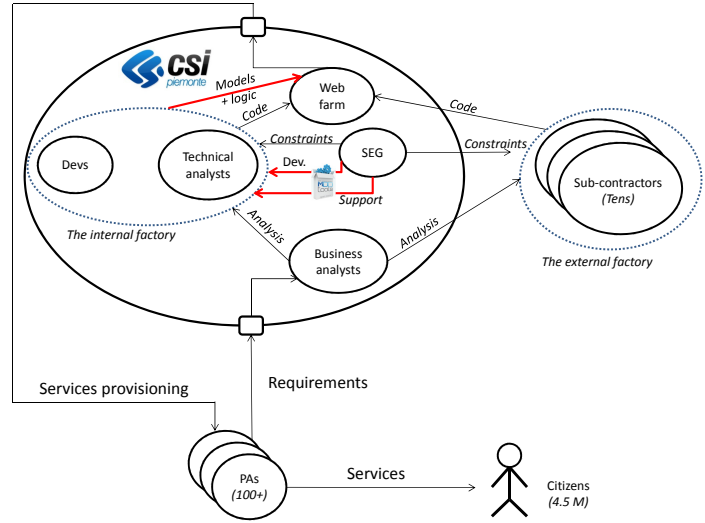}
    \caption{Ecosystem during the Investment era}
    \label{fig:ecoDuringTransition}
\end{figure*}

%\subsubsection{Description}

Given the technical viability of the MDD approach has been verified in the previous era, %at this stage the company realized it required a customized RAD tool and the management became aware of the success of the solution developed in-house for the implementation of services. %Figure \ref{fig:phase2} shows the actor involved in this phase.
the company management launched an internal project for the development of a suite of model-driven tools (named {\em MDD-Tools}). The goal of this project was both to improve and consolidate \emph{servicegen} but also to build the tools required to develop web applications presentation layer (\emph{guigen}) and data access layer (\emph{datagen}).

%\FED{Eliminati i riferimenti a Bandi per cui avrebbe senso cambiare i nomi dei progetti successivi...}
%\MTK{non sono sicuro sia politicamente consigliabile raccontarla cos\'i}
A significant investment in the MDD-Tools project was contributed as part of a project ($\alpha$). %Project $\alpha$ had not as its primary goal the realization of the MDD-Tools but it was the driver. On one hand it had to partially sustained the costs of the development of the tools, on the other hand those tools were designed taking into account the requirements of this particular project. 
%A significant portion of the the MDD-Tools development was conducted in supporting a specific project (named \MTK{\emph{Bandi}} project $\alpha$). The project clearly had not as its primary goal the realization of the MDD-Tools but it acted as the driver. On one hand it played as the testbed for the tools, on the other hand those tools were designed taking into account its requirements. 
%
After project $\alpha$ a set of pilot projects were conducted supporting with concrete evidence further decisions; the essential productivity features are presented in Table \ref{tab:productivity}. The company measures regularly the number of function points of each process using an approach based on IFPUG. 

All these projects were completed in time and reached their goals. The projects had good productivity (compared to the usual productivity measured at the same company), even better than the expectations; this is was particularly welcome because many of the developers involved had not prior experience with MDD. Circa 20 developers learnt how to use the MDD tools, and at the end of 2009 some teams were able to utilize MDD Tools without the direct involvement of the authors of the tools. %Overall the feeling of designers was good.

%The MDD-tools were evolved during the first pilot project. To be complete the project had to diverge from the pure use of the tools, overriding some of the generated code. This pragmatic approach was considered tolerable by the management in this particular case because it was the first relevant project using a full MDD approach. Nevertheless it caused difficulties in the maintenance phase and the project never adopted newer versions of the MDD tools and was forced to keep the old, original version. Considering the project bore a part of the effort for the development of MDD-Tools (which were useful in many other projects, of course) still some of the developers considered positive the impact on productivity of the MDD-Tools considering the benefit on that single project. Yet, some other developers disagreed.

%\FED{Dobbiamo rappresentare questo cambiamento nel diagramma?} \MTK{un sottoinsieme di SEG?}
A development group (within the SEG) was created and put in charge of developing the \emph{MDD-Tools} project. As a consequence, a rigorous development process for the tools was put in place with formalized mechanisms for versioning, deployment, and issue tracking.

 %At this stage the material available was quite limited. The tools started to be used by some early adopters which provided feedback. That feedback drove the successive development.

%In this phase a limited internal evangelization for the adoption of MDD-tools started.

 %In the same time there was also a change in the internal organization of the company. \FED{Questo non \'e rappresentato nel diagramma} \MTK{ lo tagliamo?} In particular the group in charge of developing the templates for web pages was dismantled. As consequence a tool for the generation of web applications presentation layer -- \emph{guigen} -- turned to be extremely useful in reducing the discomfort stemming from such reorganization.

%\MTK{In figura si parla di Analysts ma nel testo non c'\'e traccia�}
%\FED{Mauro, Paolo: cosa potremmo dire sul ruolo degli Analisti in questa fase?}

From pilot projects two problems arose: first, the need for documentation became more and more evident, then the initial reception from technical analysts was not positive. Technical analysts did not have prior experience with the tools and the solutions they proposed were sometimes not implementable with tools as described by them. Through discussions with developers it was possible to find ways to implement variants of these functionalities in a satisfactory way. However analysts were somehow reluctant to embrace the change, probably because they initially perceived the tools as limiting.

%\subsubsection{Results / Ecosystem}

Figure \ref{fig:ecoDuringTransition} shows the ecosystem during this era. The MDD-tools were provided to the software factory, together with basic face-to-face support. Only a few pilot projects were developed using the MDD-tools, they coexisted with projects developed using the traditional (non model-driven) approach.
A fundamental change in the ecosystem is that a new type of artifact started to be exchanged: models. Instead of providing the final code, the model was provided with application-specific logic -- in protected regions -- enabling the generation and re-generation of the final application.

%\MTK{It would be nice to have the numbers concerning project Bandi}

%\FED{They should be compared to results without MDD-Tools}

%Results report that:
%\begin{itemize}
%\item all the projects were completed in time and reached the goal set,
%\item the projects had good productivity, even better than the expectations. This is particularly good because there was not prior experience with the MDD tools,
%\item 20 persons learnt how to use the MDD tools,
%\item at the end of 2009 some teams were able to utilize MDD Tools without the direct involvement of the authors of the tools,
%\item the feeling of designers was good.
%\end{itemize}

\begin{table}[tb]
\centering{
\begin{small}
\begin{tabular}{lrllrr}
\hline
		 &    & \multicolumn{2}{c}{Duration} &  \multicolumn{2}{c}{Productivity} \\
Project & Size & From & To & All & Devel. \\
\hline
\multicolumn{2}{l}{Baseline}   &      &    & 15 & 30 \\
\hline
%Bandi $\alpha$ & 1600 &  ?? & ?? & ?? & ?? \\[2pt]
$\beta$ & 1091 & Oct. 2009 & Feb. 2010 & 32 & 80 \\[2pt]
$\gamma$ & 345  & Aug. 2009 & Dec. 2009 & -  & 48 \\[2pt]
$\delta$ & 307  & Sep. 2009 & Oct. 2009& 39 & 89 \\
\hline
\end{tabular}
\end{small}
}
\caption{Productivity of pilot projects vs. baseline (function points)}
\label{tab:productivity}
\end{table}

In general the results were considered positive and the organization gain the confidence for a larger adoption of the MDD Tools: during this era the MDD approach grew into a strategic asset for the company.% while the management became convinced of its importance. In particular the positive results of the first projects helped to win the initial resistances.

%\subsection{Phase 3: Maturity}
\subsection{Maturity Era}

\hfill {\bf Period:} \emph{between January 2010 and December 2012.} \vspace{1em}

\begin{figure*}[tb]
    \centering
    \includegraphics[width=.75\textwidth]{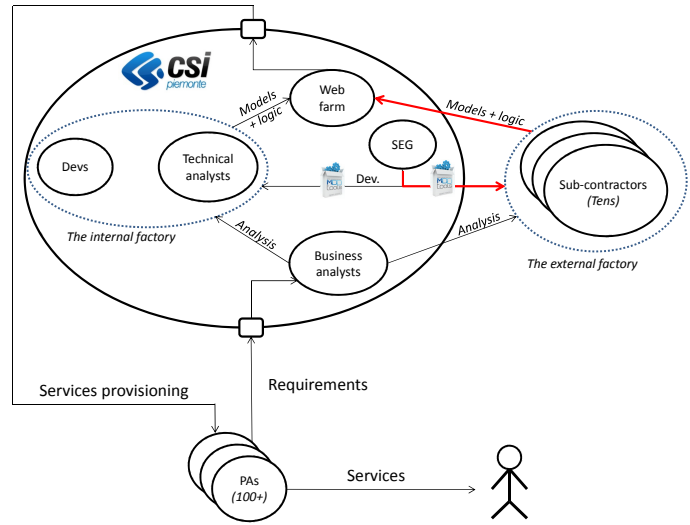}
    \caption{Ecosystem during the Maturity era}
    \label{fig:ecoMaturity}
\end{figure*}

%\begin{figure}[tb]
%    \centering
%    \includegraphics[width=0.6\textwidth]{images/UML_Diagrams/Phase3_composed.png}
%    \caption{Phase 3}
%    \label{fig:phase3}
%\end{figure}

%\subsubsection{Description}

In this era the MDD approach reaches maturity. %Positive results from prior phases justified new investments. 
Its usage was institutionalized and so it impacted more deeply the organization and started to spread to subcontractors, hence to personnel not directly employed at the company or consulting at the company offices. %several more actors.% as shown in figure \ref{fig:phase3}

%\TODO{Add the support group in the second figure}
Support was offered in a more structured way (see Sect. \ref{sec:motifSupport} for details).
%First level support for MDD-Tools was sub-contracted to an external company, while the team developing MDD-Tools remained in charge for offering second level support. This choice was motivated by the need of the development team to save time and focus on development. 
In addition, more and more developers, who already gained experience with the MDD-tools, were able to coach novice adopters. %A Q\&A knowledge base was also deployed: after a few months it contained more than 100 answered questions.

Both \emph{guigen} and \emph{servicegen} were evolved as follows:

\begin{itemize}

\item {\bf guigen:} the code generated was checked to be XHTML compliant and standardized. The support for skins was implemented:
the mechanism permitted to customize the appearance of generated applications, making easier to adapt to different PA web portals and to other kinds of customers. A specific group for the development of skins was created at CSI.
 %In previous projects adaptations of the generator on a per-portal base were done (CSI developed many applications for different PA institutional web portals). The capability to customize the appearance of each application was given by \emph{skins} based on CSS. 
The tool was evolved to support rich user interfaces, e.g. including GIS maps. A considerable effort was put to guarantee that MDD-Tools generated code compliant with the Italian regulation for accessibility which is particularly stringent for portals for the PA\footnote{See Italian law nr. 4 of the 9th of January 2004, and successive modifications}. %Finally it was added support for the portlet container Liferay\footnote{http://www.liferay.com/}.

\item {\bf servicegen:} a new generator was added to target the web services framework named Apache CXF\footnote{http://cxf.apache.org/}.
%\MTK{Apache CXF? Provide some more details}. 
%\FED{Mauro, Paolo: per CXF intendiamo Apache CXF? Riusciamo ad aggiungere più dettagli su servicen?}
Servicegen permits to define web services and their orchestration. It let the developers model the interface, user groups, authentication rules and so on. %The tool generates not only the implementation code but also stubs of unit tests, the necessary default configuration files and supporting scripts (for example the ant script to build the project). 
%\FED{Ho fatto un piccolo riassunto prendendo da http://mdd.csipiemonte.it/cms/component/docman/doc\_download/7-mdd-getting-started.html serve integrare o va bene così?}

\end{itemize}

The quality of the generated code was assessed by means of static analysis techniques. In particular the SONAR\footnote{http://www.sonarsource.org} tools was used and the generator were modified to match internally defined quality thresholds (see Sect. \ref{sec:motifQuality}). 

%An experimentation about version control repositories specific for models were conducted, with negative results. \FED{Parlare del fatto che si iniziano a indagare tematiche a corredo che sono complesse e costose, mano a mano che si adotta sono richiesti cambiamenti in ogni ambito della catena di sviluppo, il cambio di paradigma è pervasivo. L'esperimento fallisce causa mancanza documentazione e supporto tecnologia open-source}

The project was released under an open-source license, the EUPL\footnote{http://en.wikipedia.org/wiki/European\_Union\_Public\_Licence}.
Opening the project required first the legal team of CSI to review the different candidate licenses and pick up the most suitable one. On the technical side the tools needed to be reviewed to remove elements which were too much "CSI-specific". For example particular services were adopted to implement authentication and authorization of web portals developed by CSI, this aspect and others were modified to be more customizable.

The first efforts to make subcontracting companies to adopt the tool had a limited success: companies were reluctant to invest in a tool which was still considered too much CSI-specific. In particular those companies felt that the investment in training the people for the MDD-tools platform was not enough rewarding.

Anyway, during this era, the internal usage of the tool use grew. At the end of year 2011 more than 200 services were developed using MDD-Tools and more than 70 developers were able to use them autonomously. The growth in the number of users brought upfront the necessity for training, documentation and technical support (see Sect. \ref{sec:motifSupport}).

Training was provided by the team who developed the MDD tools through both scheduled internal courses and coaching during the kickstart era of the projects.
Finally the documentation was completed and enriched with tutorials, screencasts, and thematic guides.

A communication problem between analysts and developers already emerged during pilot projects; analysts did not understand the nature of the MDD-tools. This brought two types of problem: not only the requirements did not leverage the capabilities of MDD-tools, but also sometimes the requirements turned out to be not practically implementable using the MDD-tools.
To solve this issue a showcase project showing the nature of single components was developed. Using it analysts could 
learn the aspect and the utility of each one of them and started to design solutions referring to these building blocks.

%\FED{Mauro, Paolo: dovremmo spiegare bene il ruolo dello skin development team. E un ruolo nuovo che è possibile "fattorizzare" grazie a MDD?}

%At this stage the development rules previously described through documentation are now somewhat encoded in the MDD-Tools and the ecosystem consequently change as it can be seen in Figure \ref{fig:ecoMaturity}. 

%\MTK{still some repetitions below�}
%\subsubsection{Results / Ecosystem}

The configuration of the ecosystem at this stage is presented in figure \ref{fig:ecoMaturity}.
MDD-tools are not provided only to the internal factory but also to external sub-contractors, which become part of the new MDD-centered community. The sub-contractors were convinced about the cost-effectiveness of investing in MDD-tools skill acquisition with the promise of a sustained flow of jobs (see motif RoI for adopters in Sect. \ref{sec:RoI}).
At this stage the development rules previously described through documentation are mostly encoded in the MDD-Tools (see automatic enforcement motif in Sect. \ref{sec:motifEnforcement}).

The development is performed using MDD-tools in the whole ecosystem, so the deployment to the web farm is performed by providing the models and the addition logic (in the form of code within protected regions).

%At this point the company did not see anymore the MDD approach as a risk. It was convinced by positive results on many different projects. 

%Organizational issues had an effect on how the MDD tools are seen. The company decided to rely more and more on external suppliers, diminishing the percentage development realized internally therefore question arises about how the tools match this approach which was initially seen as well suited for internal development.

%\FED{Detto sotto, rimuovo: Finally the global crises affected also the possibility of performing new investments in the tool suite. While investments could not continue at the same pace as before it was clear that the users need to be supported.}

%In this period some developers started to act as evangelists in different business units.

%There were minor resistances do to the fact that particular functionalities were difficult to realize with the tool suite but in general the trade off with the increased robustness, quality and efficiency of the development process were seen as satisfying.

%\begin{figure}[tb]
%    \centering
%    \includegraphics[width=0.6\textwidth]{images/UML_Diagrams/Phase4_composed.png}
%    \caption{Phase 4}
%    \label{fig:phase4}
%\end{figure}

\FED{Draft}
Near the end of the community approach a questionnaire was distributed to some of the sub-contractors of the company to evaluate the knowledge of the tool and plan the transition to the community era. Considering that the sample was not built according to designed schema we can not assume absolute representatitivity, however 25 companies were involved. Most of the participants declared interest in learning more about the solutions as users, and a relevant part showed interested also in learning about the internals of the MDD-Tools.

%\subsection{Phase 4: the community approach}
\subsection{Community era}
\label{sec:phase3}
\hfill {\bf Period:} \emph{since January 2013.} \vspace{1em}

In this era the economic conjuncture forced CSI to reduce the investments on further development of the MDD-Tools. Nevertheless the company was aware of the necessity of guaranteeing support for existing and new users. At this stage the approach was already largely adopted by CSI and several ongoing projects were using the MDD-Tools. The company decided to focus long term efforts in fostering a stronger involvement of the community in the development of the tools.

In particular the development is not anymore performed by one single central unit; a reorganization deployed a new structure where a single person (\emph{the product leader}) is responsible for managing MDD-tools evolution by receiving contributions from several different business units and teams. While previously some users started naturally to coach and evangelize about MDD-tools this role is now officialized by the company. Moreover those active users are asked to contribute to the development of the tools itself. This change should lead to a tailorization to the needs of the different business units, which will have more control on the evolution of the suite.

Being the tool open-source and given that many subcontractors gained experience using it, it started to be used for projects not involving CSI. Unfortunately precise data on the number of these projects is not available.

\begin{figure*}[tb]
    \centering
    \includegraphics[width=.75\textwidth]{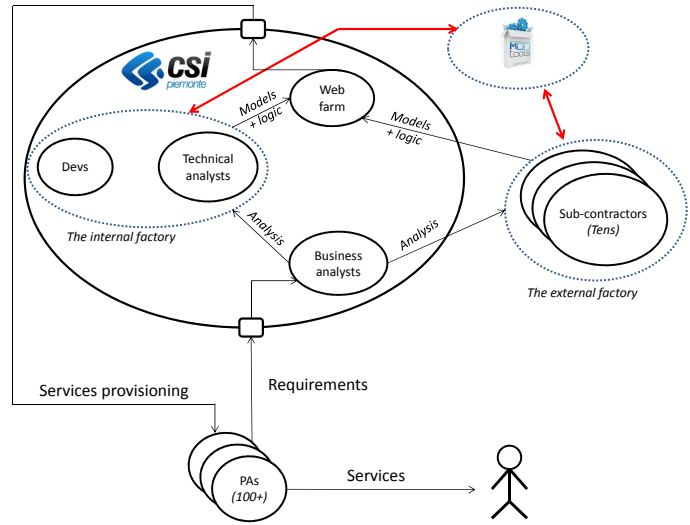}
    \caption{Ecosystem during the Community era}
    \label{fig:ecoCommunity}
\end{figure*}

%\subsubsection{Results / Ecosystem}

The company invested in a course for twelve developers of the community to improve the ability to modify the toolset itself and facilitate new contributions.

%\FED{Draft}
%MTk: rivisto.
In this period different business units contribute to the tools according to internal needs, emerging during the development of specific applications. Contributions are shared and redistributed to the community. A few contributors are starting to solve minor bugs or provide small updates even not directly related to their work, showing ownership of the toolset. For example, a contributor developed a bundle of the toolchain for Linux, while previously CSI developed bundles for Windows and Mac OS X.

Until now the community is mostly self-governed. While each user of the tool is now able to develop functionalities without the burden of synchronization and amministrative overhead on the other side there is the possibility of effort duplication or evolving the tools in different directions.
The most active members of the community start to discuss using forums, wikis and a website set-up by CSI; recently some of them are considering the possibility of producing together a new cartridge (a plugin of the generator) to support the bootstrap CSS template\footnote{\url{http://getbootstrap.com/css/}}. The cartridge previously developed by CSI were designed for the main portals of the customer PAs where the main focus was on accessibility. The adoption of the toolset in a different context requires some adaptation, mainly concerning the layout that should be rendered more responsive and ``appealing''.
%while a person is deputed to the evengalism the necessity of a board to synchronize the development is emerging. 
%Currently the company is considering to organize regular meetings between software architectes from different business units to plan the development

 %due to the necessity of require the start of an expliciti development project to the SIG on the other side there is the possibility of effort duplication or evolving the tools in different directions.

%Recently the community is discussing of producing together a new cartridge (a plugin of the generator) to support the bootstrap css template\footnote{\url{http://getbootstrap.com/css/}}. The cartridge developed by CSI were designed for the main portals of the PA, while usages of the toolset in different contexts require an adaptation of the layout more focused on the responsiveness and "design appeal" instead that on accessibility, as it was the focus of the software developed for the PA.

The ecosystem at this stage is depicted in figure \ref{fig:ecoCommunity}. The MDD-tools open-source project has now become a new stand-alone entity in the ecosystem. The developers in the ecosystem now focus on contributing new features of the MDD-tools, they provide new code to the OSS project. In practice we observe a transition from a centralized development typical of closed-source environment to a distributed development typical of an open-source environment (see Sect. \ref{sec:DistributedDevelopment}).

%\section{Reception and results}
%\label{sec:results}
%\input{results}

\section{Motifs}
\label{sec:motifs}
\begin{table*}[tb]
\centering
\begin{tabular}{lccl}
\hline
\multirow{2}{*}{{\bf Motif}} & \multicolumn{2}{c}{{\bf Effects on}} & \multirow{2}{*}{{\bf Era}} \\
& {\bf Efficiency} & {\bf Diffusion} &  \\ \hline 

Incremental adoption &  & \checkmark & Informal \\
Toolsmithing & \checkmark &  & Assessment \\
Integration & \checkmark & \checkmark & Investment \\
Support &  & \checkmark & Maturity \\
Automatic enforcement & \checkmark &  & Maturity \\
Generated code quality & \checkmark &  & Maturity \\
RoI for adopters &  & \checkmark & Maturity  \\
Distributed development &  & \checkmark & Community \\
\hline

\end{tabular}
\caption{Motifs with main effect and era of appearance}
\label{tab:motifEffects}
\end{table*}

The way the ecosystem was shaped by the introduction of the MDD-tools and the way CSI managed to successfully spread over the new paradigm -- first internally to the organization, then to the whole ecosystem -- are main features we observed in our investigation.

%The evolution of MDD-tools across the six years of their history represent a reservoir of software engineering practices.
Through different forms of interactions that took place in the last two years -- in the context of a collaboration between CSI and Politecnico di Torino -- we elicited several of such practices. %\TODO{Check if coherent with methodology}

The are first of all an abstraction of our interpretation of the evolution of the ecosystem and secondly they could turn out to be a reservoir of potentially useful software engineering practices.

We dubbed {\em ``motif''} the most relevant among the elicited software engineering practices. We use the term {\em motif} because they cannot be considered proper organization \cite{Coplien1998} or design patterns \cite{Gamma1995}, both because they are not proved\footnote{We stress here that our approach is mainly interpretive, we cannot claim a positivist nature for our results, though they could be used as the basis for further a positivist research} and they cannot always be formulated as solutions.

The motifs we present here, we believe, are peculiar of the evolution of an ecosystem when a new development tool or technology -- produced and used by the different members of the ecosystem -- is introduced.
Their applications -- in different eras of the ecosystem history -- turned out to be success factors for the favorable evolution and growth of the ecosystem.

We describe each motif using the so called "Canonical form"\footnote{\url{http://c2.com/cgi/wiki?CanonicalForm}} for pattern description, and we report how the motif was applied in the case under study.
%\MTK{or we use the simplified version?}

Each motif was applied mostly in one specific era as shown in Table \ref{tab:motifEffects}. They affected two main aspects: the efficiency of the software development and the diffusion of the MDD approach within the ecosystem.

\subsection{Incremental adoption}
\label{sec:motifIncrementalAdoption}

\begin{table*}[tb]
\begin{scriptsize}
\begin{tabular}{|>{\raggedleft}p{0.11\textwidth}p{0.83\textwidth}|}
\hline
{\bf Name}		& Incremental adoption \\[2mm]
%\hline
{\bf Problem} & There is resistance to the diffusion of the (development) technology \\[2mm]
%\hline
{\bf Context} & The diffusion is in its initial phase, from the initial proposers' group 
				  to a still small group of potential new adopters \\[2mm]
%\hline
{\bf Forces}  &  Tendency to maintain work habits \hfill \  \linebreak 
				  Skepticism about benefits \hfill \  \linebreak
				  Fear of possible difficulties\\[2mm]
%\hline
{\bf Solution}& Conduct the adoption of the technology incrementally and on a voluntary base \\[2mm]
%\hline
{\bf Resulting \linebreak context} & Word of mouth dissipates fears and skepticism, 
							required knowledge is easily available from the neighbours \\[2mm]
%\hline
{\bf Rationale} & This approach allows for a spontaneous and persuaded adoption
					 and facilitate the natural emergence of champions \\
\hline
\end{tabular}
\end{scriptsize}
\caption{Motif summary: Incremental adoption}
\label{tab:motifIncrementalAdoption}
\end{table*}

%The progressive and voluntary diffusion of the MDD approach allows a spontaneous and persuaded adoption and the natural emergence of champions.
%The adoption of MDD in CSI was incremental. 
Before the initiative started as a formalized task by SEG in 2008, a few tools were already used to perform code generation. Single developers were using tools like XDoclet\footnote{\url{http://xdoclet.codehaus.org/}} or FreeMarker\footnote{\url{http://freemarker.sourceforge.net/}}. The former allows generating code from annotations\footnote{\url{http://en.wikipedia.org/wiki/Java\_annotation}} inserted in Java modules, while the latter is a template-based code generator. Because of this personal experimentation some of the developers where familiar with code generation and had experienced its benefits on real projects.

When the SEG introduced \emph{csiskelgen} (i) it was an incremental evolution in respect to previous approaches, and (ii) the adoption was conducted on a voluntary basis. During initial phases developers were free to adopt it or not for new projects, without pressure from project managers or executives. Later the adoption of the tool and its successors spread naturally into different business units thanks to word of mouth and evangelization spontaneously conducted by satisfied users. When the management started to encourage a more structured adoption of MDD-Tools many developers had already either used them personally or heard of success-stories from colleagues.

According to Rogers \cite{Rogers2003} in this case a {\em collective innovation decision} is taken in favor of the adoption of MDD.
This situation contrasts with scenarios where the adoption of MDD is forced by the management and the developers of the organization have no prior experience with it. In those cases resistances to adoption are probable \cite{harrison1997executive}.
The gradual approach is also suggested in \cite{HutchinsonPhd}.

%\begin{table}[tb]
%\begin{scriptsize}
%\begin{tabular}{|>{\raggedleft}p{0.11\textwidth}p{0.87\textwidth}|}
%\hline\noalign{\smallskip}
%\multicolumn{2}{|l|}{\bf Incremental adoption} \\[2mm]
%%\hline
%{\bf Problem} & There is resistance to the diffusion of the (development) technology, 
%					since developers tend to keep their habits, are skeptic
%					about the benefits, and wish to avoid the effort of
%					getting the required new knowledge. \\[2mm]
%%\hline
%{\bf Context} & The diffusion is in its initial phase, from the initial proposers' group 
%				  to a still small group of potential new adopters \\[2mm]
%%\hline
%{\bf Solution}& Conduct the adoption of the technology incrementally and on a voluntary base.
%This approach allows for a spontaneous and persuaded adoption and facilitate the natural emergence of champions.
%Word of mouth dissipates fears and skepticism, 
%							required knowledge is easily available from the neighbor
% \\[2mm]
%\hline
%\end{tabular}
%\end{scriptsize}
%\caption{Motif summary: incremental adoption}
%\label{tab:motifIncrementalAdoption_}
%\end{table}

%\subsection{Autonomus initiative from developers}

%Manager do not care and developers are free to realize their own tools if a large investment is not required. 
%In this case there was a prior experience with dedicated tools for the generation of code for some of the layers of the application (for example for the data access layer: think about tools like iBatis and hibernate).

\subsection{Toolsmithing}
\label{sec:motifToolsmithing}

There are several tools in the marketplace that support the development of business application using an MDD approach. During the Assessment Era the SEG carried on an evaluation of the most significant ones.

The result of the evaluation, was that in a large software factory as the one present in CSI the best solution possible seems the development of custom tool-chain and structure which could be shaped and adapted to the organization's need. Alternatives based on the adoption of packaged solution were evaluated and discarded in the case of CSI. %\FED{Tolto la parola DSL, anche Webratio usa un DSL, il punto qui e' che io controllo la soluzione e posso modificarla a mio piacimento perche' ne controllo gli internal. Come compagnia che sviluppa software, independentemente dalla dimensione, voglio avere il pieno controllo di questi tool in cui embeddo le mie competenze: come dire, se sono FIAT non do in outsourcing lo sviluppo del motore}

%\TODO{BOZZA}
%MTk: OK rivisto 
Toolsmithing, and in particular the definition of specialized editors seems to be rarely implemented in MDD solutions, with micro companies slightly more inclined to consider it in respect to large companies \cite{tomassetti2012_JSS}. This is probably due to the fact that more complex processes are typically adopted in the latter, which could be more reluctant to develop their own tools and perform the necessary complimentary steps for a successful deployment (e.g. revise the processes, train the developers).

It is true that a significant effort is required for the development of appropriate technological knowledge to enable such an approach, though this study suggest that practitioners working in software-intensive companies can consider this possibility as a viable option. An industrial survey confirms the role of toolsmithing in improving the chance to achieve improved flexibility, productivity, reactivity to changes and platform independence \cite{Torchiano2013}.
To partially reduce the development cost we also suggest to base custom MDD solutions on enabling platforms like EMF or commercial alternatives (e.g., MetaEdit+).

\begin{table*}[tb]
\begin{scriptsize}
\begin{tabular}{|>{\raggedleft}p{0.11\textwidth}p{0.87\textwidth}|}
\hline
{\bf Name}		& Toolsmithing \\[2mm]
%\hline
{\bf Problem} & The needs of the company are not met by existing commercial tools \\[2mm]
%\hline
{\bf Context} & The technological evolution compels a company to adopt new tools \\[2mm]
%\hline
{\bf Forces}  &  Existing tools impose their approach \hfill \  \linebreak
				  The company has a set of constraints that contrast with the tool's approach \\[2mm]
%\hline
{\bf Solution}& Develop the tool in house  \\[2mm]
%\hline
{\bf Resulting \linebreak context} & The constraints are satisfied by the new tool \\[2mm]
%\hline
{\bf Rationale} & Building the tool allows for a perfect customization to the technical and process needs of the company \\
\hline
\end{tabular}
\end{scriptsize}
\caption{Motif summary: Toolsmithing}
\label{tab:motifToolsmithing}
\end{table*}

\subsection{Integration}
\label{sec:motifIntegration}

%As the number of adopters grows, a more than linear growth in the effort devoted to documentation and integration with customary tools is required in order to support them.

The initial tool (\emph{csiskelgen}) was not integrated with the IDE and it was quite unpolished, still it was perceived as valuable from developers.

IDE integration turned out not being critical to promote the first adoption of the tool but it had to be addressed to permit to a larger portion of the users to adopt it. 

This turned out to be a successful action in the Investment Era (see section \ref{sec:phase2}.

The selected technological platform (Eclipse EMF) is well suited to build plug-ins for the Eclipse IDE that is typically used by the developers in the ecosystem. The integration with the platform permits to benefit of accessory facilities (like plugins for supporting version-control systems) with which the user-base is already familiar. Integration can be seen as a way of mitigate risks and to leverage existing investments \cite{Selic2003} both on tools development or acquisition and effort spent on skill development.

Balasubramanian et al. reported missing integration as a problem making difficult to obtain a complexive view of the system \cite{Balasubramanian2006}. We agree with this finding and we suggest to practitioners to consider tools integration when planning the diffusion of technological changes to a large number of different actors. 

\begin{table*}[tb]
\begin{scriptsize}
\begin{tabular}{|>{\raggedleft}p{0.11\textwidth}p{0.87\textwidth}|}
\hline
{\bf Name}		& Integration \\[2mm]
%\hline
{\bf Problem} & New adopters have problems in using the technology that doesn't fit their usual workflow \\[2mm]
%\hline
{\bf Context} & The technology starts to be adopted by a larger group not including only enthusiasts \\[2mm]
%\hline
{\bf Forces}  &  Developers tend to stick to their workflow \hfill \  \linebreak
				 Tools implicitly assume a given organization of work \\[2mm]
%\hline
{\bf Solution}&  Integrate the tools into the commonly used development environment (e.g. as plug-ins)\\[2mm]
%\hline
{\bf Resulting \linebreak context} & No significant change to the workflow is required to use the tools \\[2mm]
%\hline
{\bf Rationale} & A least effort solution that allows using the usual IDE without heavily affecting the workflow 
					 can potentially overcome the resistance from developers\\
\hline
\end{tabular}
\end{scriptsize}
\caption{Motif summary: Integration}
\label{tab:motifIntegration}
\end{table*}

\subsection{Support}
\label{sec:motifSupport}

\begin{table*}[tb]
\begin{scriptsize}
\begin{tabular}{|>{\raggedleft}p{0.11\textwidth}p{0.87\textwidth}|}
\hline
{\bf Name}		& Support \\[2mm]
%\hline
{\bf Problem} & New adopters have problems in using the technology: they need to train, learn and acquire the specific skills \\[2mm]
%\hline
{\bf Context} & Scaling up from a small (voluntary) group, through management commitment \\[2mm]
%\hline
{\bf Forces}  & Need for documentation and support \hfill \  \linebreak
                Community used direct personal contacts \\[2mm]
%\hline
{\bf Solution}& Plan of an heavy effort investment in documentation, building a support group \\[2mm]
%\hline
{\bf Resulting \linebreak context} & Information is available through the official documentation 
									 or resorting to support group \\[2mm]
%\hline
{\bf Rationale} & When a certain size threshold for the adopted group is reached, nearby colleagues are not anymore available or sufficient for training and problem solving therefore an institutionalized approach to knowledge sharing must be implemented\\
\hline
\end{tabular}
\end{scriptsize}
\caption{Motif summary: Support}
\label{tab:motifSupport}
\end{table*}

%As the number of adopters grows, a more than linear growth in the effort devoted to documentation and integration with customary tools is required in order to support them.

While for popular and widespread development technologies the knowledge is normally freely available in the web, for tools used only within a certain ecosystem (like MDD-Tools) specific support have to be provided from within the ecosystem itself.

In this case support is needed both during development and maintenance. During development programmers need clarifications about the modeling language and advices on the best practices, during maintenance help is needed for tuning the application and to conduct bug fixing. 

Initially the documentation was very limited and there was not a group in charge of offering support, this was due to the nature of the initial incremental and voluntary adoption process (see Sect. \ref{sec:motifIncrementalAdoption} above). At that time users just turned their question to the initial contributors or other developers in the same team.

%\FED{Draft}
%MTk Revised
The transition from in-house usage of the MDD solution to the ecosystem configuration required an adequate management of knowledge. In this respect the situation is similar to what happens with off-shoring: knowledge which existed internally within an organization's boundaries is moved to and exchanged with external organizations \cite{Bahli2005}. In the CSI case the knowledge considered includes both technical expertise and know-how about processes. 

The importance of knowledge transfer is fundamental to properly perform this step and both explicit and tacit knowledge should be considered \cite{Swartbooi2010}. In our case-study the explicit knowledge was transmitted through documentation while the tacit knowledge was accessible through the support team and with dedicated training-on-the-job activities.

When the ecosystem grew at a fast pace during the Maturity Era (see section \ref{sec:phase3}), such approach became infeasible, thus leading to different strategies to guaranteeing support. First the SEG offered both the first and second level support, as the workload kept growing  the first level support had to sub-contracted to an external company. 
%In addition CSI provided support through a unit which replied directly to questions from the users. 
A Q{\&}A system (similar to Stackoverflow\footnote{http://stackoverflow.com/}) was also implemented when the same questions started to appear over and over. The system was made available to all the developers working in the ecosystem. 

It is important to emphasize how value can be provided to a small initial circle of users with a limited investment while the leap out of the circle -- i.e. delivering value to a wider population of developer -- require a much larger investment.

Though, at a later stage, as the tools are more and more adopted the number of advanced users able to offer some support to colleagues grows.

It is important for practitioners to consider both explicit and tacit knowledge and be ready to perform the necessary investments, if they want to fully exploit the benefits of a technological transition.

% MOVED TO DISCUSSION BECAUSE it is difficult to propose as a motif
%\subsection{Support}

%There are several dimensions that need to be balance in providing support to the adopters: staff vs. knowledge base applications, centralized vs. distributed, responsibility on developers vs. support team.

%While for popular and widespread development technologies the knowledge is normally freely available in the web, for tools used only within a certain ecosystem (like MDD-Tools) specific support have to be provided from within the ecosystem.
%Support needs to be guaranteed both during development and during maintenance. During development programmers need clarifications about the modeling language and advices on the best practices, during maintenance help is needed for tuning the application and to conduct bug fixing. 

%Offering support in a centralized way has some negative aspects: mainly the person offering support typically does not know the project and is not aware of it peculiarities. As second point he does not feel involved in the project. In addition, it is possible that the central unit offering support gets to assume too many responsibilities, while single business units would be not be considered fully responsible for the correct implementation of their projects.

%
%CSI faced some bureaucratic limitations in providing teachers to sub-contractors because this particular activity was not included in its statute.

\subsection{Automatic enforcement}
%\FED{Lo chiamarei Formalization of development rules or Automatic enforcment}
\label{sec:motifEnforcement}

\begin{table*}[tb]
\begin{scriptsize}
\begin{tabular}{|>{\raggedleft}p{0.11\textwidth}p{0.87\textwidth}|}
\hline
{\bf Name}		& Automatic enforcement \\[2mm]
%\hline
{\bf Problem} & It is difficult to make several different providers (sub-contractors) to comply with a given set of standards \\[2mm]
%\hline
{\bf Context} & The developed software must comply to legal constraints (e.g., on usability) and technical constraints deriving from the web farm platform
 \\[2mm]
%\hline
{\bf Forces}  &  Conformance to (usability/integrability/standards) rules is required \hfill \  \linebreak
				Target platform may vary \hfill \  \linebreak
				Developers and sub-contractors tend to adopt most familiar/cheap technologies\\[2mm]
%\hline
{\bf Solution}& Encode the rules in the code generator \\[2mm]
%\hline
{\bf Resulting \linebreak context} & Conformance is guaranteed by the tool 
 \\[2mm]
%\hline
{\bf Rationale} & Instead of imposing directly the constraints, e.g. through heavy rules and standards books, the constraints are encoded in the tool and the conformance is automatically -- and mostly transparently -- ensured by the everyday working tools, e.g. by generating conforming code
\\
\hline
\end{tabular}
\end{scriptsize}
\caption{Motif summary: Automatic enforcement}
\label{tab:motifAutomaticEnforcement}
\end{table*}

%\FED{Caption e label del pattern Governance of applications life cycle sono da rivedere}

%By focusing the whole approach -- including the artifacts produced by sub-contractors -- on MDD-tools a strict governance of the whole application life cycle is possible.
%The governance of MDD applications presents several advantages in respect to the traditional forms of development.

The control of several different aspects of the applications is fundamental. For instance from a functional perspective the applications must be hosted by the CSI web farm and be integrated with common platform modules, e.g. authentication. From a non-functional point of view, web portals developed for the PA must conform to a set of accessibility standards.

MDD-tools make possible the governance of every aspect of the applications life cycle. Sub-contractors could in principle provide just the models -- accompanied by application-specific logic that need to be written manually within the protected regions -- to CSI, which then autonomously generates and deploys the application. This approach permits the enforcement of standards and procedures with a limited effort.  
%While this process is not yet implemented it is something considered for the future.

Upon receiving the models, CSI has the option to check them to verify that quality requirements are met, for instance the rules for usability could be mostly automatically verified. This is made possible because an high level of abstractions is adopted (models instead of code). 
In addition it is possible to use a standardized building process to generate the code and produce the final application. Finally, the deployment also can be performed in a standardized way.

%MDD-tools make possible the governance of every aspect of the applications life cycle made possible by MDD-tools is so simplified that also the final customer (i.e., the PAs) could perform it almost autonomously\MTK{la regione ce la fa da sola?}.  
The transition to the MDD-based ecosystem with automatic enforcement during the Maturity Era (see section \ref{sec:phase3}), brought significant benefits.
This approach contrasts with the reality in other contexts -- e.g. PAs in close regions -- which experience strong difficulties in managing, installing and maintaining a plethora of interdependent applications developed with different technologies, on different platforms and with different requirements. 

The idea of enforcing architectural rules was envisioned before \cite{Mattsson2008}. Normally it is executed as a supplementary step in which architectural rules are automatically or manually verified \cite{Mattsson2010}. In this case rules were not anymore formalized explicitly, but they were embedded in the generators and the solution itself.

We suggest practitioners to consider the benefits of automatic-enforcement in managing relations across an ecosystems. By reducing the cost of evaluation and making more objective the process, the level of confidence in the relation can grow.

%Work on establishing ways to measure the quality of models should be done.

\subsection{Quality of the generated code}
\label{sec:motifQuality}

\begin{table*}[tb]
\begin{scriptsize}
\begin{tabular}{|>{\raggedleft}p{0.11\textwidth}p{0.87\textwidth}|}
\hline
{\bf Name}		& Generated code quality \\[2mm]
%\hline
{\bf Problem} & The poor quality of generated code affects the overall quality of the product\\[2mm]
%\hline
{\bf Context} & The development tool produces code, that will be shipped as part of the final product, e.g. using an MDD approach
 \\[2mm]
%\hline
{\bf Forces}  &  Quality of generated code is rarely put into question \hfill \  \linebreak
Generated code has the potential to cripple the whole application \hfill \  \linebreak
Generated code could need to be understood, e.g. to integrate, debug, troubleshoot \hfill \  \linebreak
Generated code is not intended for developers to read
\\[2mm]
%\hline
{\bf Solution}& Assess the quality of generated code and improve the generator \\[2mm]
%\hline
{\bf Resulting \linebreak context} & The final quality of the generated code is improved, which is reflected in improved overall quality of the product. In addition the understandability of the code is improved
 \\[2mm]
%\hline
{\bf Rationale} & Several quality issues may indicate bugs that otherwise would be difficult to understand, and the resulting improved quality often implies better understandability
\\
\hline
\end{tabular}
\end{scriptsize}
\caption{Motif summary: Generated code quality}
\label{tab:motifCodeQuality}
\end{table*}

%The assessment of the quality of the generated code represents a leverage to improve the quality of the whole applications pool. 

During the Maturity Era (see section \ref{sec:phase3}), the SEG performed an investigation for quality issues on the generated code and individuated a few actual bugs.
It is important to notice how an investment on the quality of the generated code could be immediately retrofitted to affect all the projects developed using the code-generation feature of the platform.

Normally the quality of generated code is not monitored; this case suggest it could make sense to do that because:
\begin{itemize}
\item the generated code could need to be integrated with custom code written by developers and not generated,
\item also if manually written code is not inserted the generated code could still need to be read and understood for debugging and troubleshooting,
\item quality problems in the generated code could lead to errors or performance issues.
\end{itemize}

Investigation on the quality of the generated code is often neglected. We suggest it useful to improve two factors: performances and code readability. For most applications the performances are simply not an issue, while for applications where performances are critical MDD tend to be rejected to retain full control of the executed code \cite{Selic2003}. Code readability of generated code is normally not much considered: the best-practices suggest that all code should be generated \cite{Kelly2008} and considered as a semi-artifact necessary just to obtain the compiled applications. We argue it is not frequently the case: in many MDD solutions a large fraction of code is generated but corner-cases are managed through hand-written code \cite{Torchiano2013} which need to be integrated with generated code. In this scenario the readability of generated code becomes important.

Investments in the quality of the generated code can be incremental and benefit not only applications currently being developed but also past ones -- by means of a regeneration of the code -- and future ones. On the contrary investments on the manually written code of one applications can benefit only the individual application being targeted.

Working on improving the quality of code, previous findings should be considered: previous research \cite{Vetro2011Findbugs, Vetro2013_SEW, Vetro2010_MSR} indicates that while a portion of warnings issued by static analysis tools are highly correlated to potential errors, most of them are not. The latter type of indicators cause an high workload corresponding to a small or null contribution to the improvement of quality. Therefore when evaluating the quality of the generated code the relevant indicators need to be identified using techniques which require competencies on the technologies involved and on statistics.

%\FED{Dici che non e' il caso di dire che gli indicatori per il codice generato potrebbero essere diversi da quelli standard?} %No, interessante ma c'� troppa carne al fuoco..
%It is not sure if all the indicators which are proved to be good for assessing the quality of code manually written do apply also in the case of inspection of generated code. It could be that only some quality (e.g., readability but not maintainability) are valuable for generated code. Future work could be done in this direction.

%\FED{Draft}

\subsection{RoI for external adopters}
\label{sec:RoI}

\begin{table*}[tb]
\begin{scriptsize}
\begin{tabular}{|>{\raggedleft}p{0.11\textwidth}p{0.87\textwidth}|}
\hline
{\bf Name}		& RoI for new adopters \\[2mm]
%\hline
{\bf Problem} & New adopters are reluctant to invest in training on new technology \\[2mm]
%\hline
{\bf Context} & The development tools requires a non-negligible effort in training and learning, while the technology is mainly used within the ecosystem
 \\[2mm]
%\hline
{\bf Forces}  &  Large upfront training investment \hfill \  \linebreak
					Limited reusability of the new skills \\[2mm]
%\hline
{\bf Solution}& Provide a medium-/long-term commitment in providing jobs \\[2mm]
%\hline
{\bf Resulting \linebreak context} & Upfront investment is repaid by long-term job series
Acceptance at company level may trigger a similar problem at individual level
 \\[2mm]
%\hline
{\bf Rationale} & The new adopters (sub-contractors) must have some form of guarantee that the investment yields a return through a long-term job agreement
\\
\hline
\end{tabular}
\end{scriptsize}
\caption{Motif summary: RoI for adopters}
\label{tab:motifRoI}
\end{table*}

%Adoption by new members of the ecosystem involve a significant investment in learning, with can be justified only if it yields a larger return in terms of work opportunities.

%CSI had the necessity to convince suppliers to support the initiative and the change affecting the whole ecosystem.

The adoption of the MDD-tools by sub-contractors required, on their side, a significant investment in acquiring the specific skills for using that platform. Such knowledge, though, could only be profitable when developing for CSI.
During the Maturity Era (see section \ref{sec:phase3}), the decision of a sub-contractor to adopt the MDD-tools therefore consists in evaluating the return yielded by such an investment. 

%In this case CSI had to gain the support of suppliers guaranteeing a certain amount of work to be sub-contracted to them. 
An affordable return on investment is possible if CSI is able to guarantee a given amount of work to be sub-contracted.
Once sub-contractors felt that there was a sufficient yield they were willing to invest in training for MDD-tools, with the possibility of using it also for other customers.

While this strategy eventually led most sub-contracting companies to decide for the adoption of MDD-tools, individual developers in such companies were sometimes still reluctant in investing too much in learning the MDD-Tools. Their fear was to reduce their ability to find a different job, because the skills acquired with MDD-Tools are relevant only inside the ecosystem of companies using them. Developers would prefer to invest in skills which provide access to a larger job-market (e.g., java programming). In this case the advantage of the business unit or the benefit of the supplier need to be put in front of the prospective of the single developer. On the other end developers using MDD-Tools slightly change their role and develop a different skill-set which could be useful to work with other higher-level DSLs. To individuals to benefit of this acquired skills the industry need to adopt them more largely.

%\subsection{Changing roles}
%\MTK{Questo \'e un po' troppo specifico }

%In CSI two kinds of analysts operate along with developers: i) technical analysts and ii) business analysts \FED{Qui servirebbe una mano per definire i ruoli dell'uno e dell'altro}.

%Technical analysts are reluctant to model because it means to provide a technical artifact which is used to produce the concrete applications therefore more responsibilities are involved, while errors in the technical documentation currently provided do not lead to too strong consequences.

%On the other end the developers do not want to just models because the skills acquired lose value.

%The business analysts instead keep collecting requirements, its role does not change.

%Modellers are a new intermediate figure. They could be seen as a designers who could benefit from some knowledge about user experience. 

%While in other realities the figures of the technical analysts and the developer could be merged in one person this is not yet the case in CSI.

\subsection{Distributed platform development}
\label{sec:DistributedDevelopment}
\begin{table*}[tb]
\begin{scriptsize}
\begin{tabular}{|>{\raggedleft}p{0.11\textwidth}p{0.87\textwidth}|}
\hline
{\bf Name}		& Distributed development \\[2mm]
%\hline
{\bf Problem} & Centralized development team can hardly identify direction for enhancing the tools \\[2mm]
%\hline
{\bf Context} & A centralized team manages the development a fairly mature technology \\[2mm]
%\hline
{\bf Forces}  &  Difficult to identify directions for evolution \hfill \  \linebreak
					Lack of knowledge about the specific \hfill \  \linebreak
					Need to evolve the technology\\[2mm]
%\hline
{\bf Solution}& Distribute the development to the individual business units that use the technology
 \\[2mm]
%\hline
{\bf Resulting \linebreak context} & The needs are address by evolution within the same context where they arise \\[2mm]
%\hline
{\bf Rationale} & The evolution is pushed and driven by specific needs, plugged into a stable and mature core
\\
\hline
\end{tabular}
\end{scriptsize}
\caption{Motif summary: Distributed development}
\label{tab:motifDistributedDevelopment}
\end{table*}

%Once the MDD-tools platform reached its maturity, the development can be distributed and only a limited effort and staff is required centrally to coordinate the evolution.

Since the MDD-tools platform reached its maturity, further evolution mainly address new needs that emerge in different domains during the development. The central team -- the SEG at CSI -- in charge of the tools' evolution encountered several difficulties in collecting, understanding and implementing the requirements coming from the development teams.

Within a large ecosystem, not only the evolution becomes critical, offering a centralized tool support shows some drawbacks: the person offering support typically does not know the project and she is not aware of its peculiarities and does not feel involved in the project. In addition, this division introduces an attribution problem \cite{Jaspars1983}: the people in central support unit tend to be considered responsible for most problems even loosely linked to the MDD-tools and they feel they are blamed too often, while developers working in the other business units tend not to considered themselves fully responsible for the correct implementation of their projects. %\TODO{Maybe some relwork on the distribution of responsibilities} 
%MTk: potrebbe andare ma a questo punto \'e tardi

In an attempt to solve the above problems the tools' evolution was later moved to a distributed schema: individual business units are in charge of developing the extensions they need. External partners are welcome as well to provide contributions. Most of the people originally working in the support unit have been relocated in the business units, while at the central level is maintained only the role of coordinating the development of the tools, to maintain a cohesive strategy and to avoid duplication. %\FED{forse possiamo definire meglio questo ruolo}

\section{Discussion}
\label{sec:discussion}
%\FED{Il concetto da esprimere qui e' perche' l'ecosistema finale e' meglio di quello iniziale.}

The whole history of the ecosystem evolution represents a continuous improvement both for the collective and the individuals from several perspectives. 
The final configuration of the ecosystem is an improvement in respect to the initial situation for a variety of reasons.
Here we summarize the changes occurred in the ecosystem and its evolution along a few relevant dimensions.

\begin{description}

\item[Productivity:] the development process is based on more productive tools, which permit to consistently reduce the development cost and provide more predictable development time. The productivity measured in the Investment era (see Table \ref{tab:productivity}) indicates an improvement factor ranging from 1.5 to 3, w.r.t. the baseline. This result is consistent with a previous survey of modeling and MDD in the Italian industry \cite{Torchiano2013}.

\item[Skills:] the ecosystem allowed many companies of the regional IT district to acquire competencies on MDD which were scarce before. As MDD-Tools are adopted by sub-contractors not for new customers, different from CSI, the competencies developed on MDD-Tools become more useful and led more practitioners to learn about MDD. Actually the lack of competencies has been reported among the top three reasons why MDD techniques are not adopted in the Italian industry \cite{Torchiano2013}; we expect the new skill will trigger a virtuous circle that will lead to a wider adoption of such techniques in the area.

\item[Control:] before the adoption of MDD-Tools, manual code inspections were necessary to verify the adherence of applications to the company development standards. After the adoption of MDD-Tools a large portion of code is automatically generated from models and therefore the generation process automatically ensure the strict adherence to those standards. In summary MDD-Tools significantly simplified the governance of the applications. Transitions to new platforms are now possible and cheaper, because the technological aspects are captured only into the code generators, instead of being spread through all the codebase. This permits to migrate applications as the implementations become obsolete, preserving the investment done by customers. For example a transition from MySQL to PostgreSQL in the CSI web farm was performed during the period of interest: thanks to the generative approach it required to adapt only the generators and it was possible to apply it to a large number of applications with a limited effort. 

In the future it could be possible for CSI to acquire directly and exclusively models which are easier to verify. Some experimentation on metrics for MDD-Tools models are already started. %\FED{diamo un'occhiata qui http://mdd.csipiemonte.it/cms/risorse/74-model-driven-metrics-overview.html }
They could be an important tool for the final customer to evaluate the quality of the applications received and to permit a fairer competition between different suppliers.

\item[Flexibility:] In a software ecosystem involving tens of companies, as the one centered on CSI, there are several specific needs and constraints, most of which percolate from the customers -- i.e. several PAs --. During the \emph{Assessment} era (see Sect. \ref{sub:phase1}) a survey and assessment of commercial packaged solutions was conducted; they were discarded because of lack of flexibility.

The MDD-tools were instead developed as custom tools and structure which could be personalized using a DSL, building upon the enabling platform of EMF. Such a solution proved itself flexible enough to accommodate all the different development needs that emerged in the ecosystem. Toolsmithing represented a success factor, confirming what emerged from a previous survey \cite{Torchiano2013}. The necessity to be able to craft their own tools seem fundamental for software intensive companies of each size, as emerged also in a case-study previously conducted by authors of this paper \cite{Tomassetti_Jucs_2013}.

%\item[Abstraction:] Companies willing to work with CSI do not have to study the development rules specific to CSI but can learn instead how to model (\FED{Ammesso che questo sia meglio}). Therefore CSI has a lower lock-in in the choice of sub-contractors.

\item[Roles:] %\MTK{ this a possible future development, or since no evidence we could remove it}
%\FED{L'aspetto di chi scriva i modelli viene fuori in diversi lavori, il fatto che MDD faccia emergere un certo tipo di figura professionale mi sembra interessante, in effetti pero' potrebbe non essere un vero e proprio miglioramento e pertanto non starci qua se non rivisto}

Traditionally two figures constitutes the bulk of the employees of the internal and the external factory: technical analysts and developers. With the switch to a MDD paradigm of development these factories need to produce mainly models and the accompanying logic (still written using general purpose languages). 

Technical analysts are reluctant to model because it means providing a technical artifact which is used to directly produce the concrete applications therefore more responsibilities are involved. While errors in the technical documentation -- e.g. word files -- currently do not lead to dire consequences. On the other end the developers do not want to just model because the related skills are less useful -- in terms of their career -- in other contexts than the knowledge e.g. of some mainstream programming language.

Currently models are written mainly by people trained as software developers. In the long run the change operated in the ecosystem with the adoption of MDD call for a redefinition of the figures involved in the software development because the skill set required for writing models using MDD-Tools seems slightly different from the one of developers and technical analysts. The optimal solution seems the definition of a new role for which at the moment there is an absolute scarcity in the ecosystem. Modellers are a new intermediate figure. They could be seen as designers who could benefit from some knowledge about user-experience design principles and usability. \TODO{Rephrase}

Given the number of personnel employed in the ecosystem it is a need that could affect the regional market job and should be considered also from the local universities.

%With thousand of IT personnel employed by CSI and su

% In CSI two kinds of analysts operate along with developers: i) business analysts and ii) technical analysts; the former elicit the business requirements by leveraging their knowledge of the domain, the latter specify the requirements with reference to the technological platform.

%On the other end the developers do not want to just models because the skills acquired lose value.
%The business analysts instead keep collecting requirements, its role does not change.
%Modellers are a new intermediate figure. They could be seen as a designers who could benefit from some knowledge about user experience. 
%While in other realities the figures of the technical analysts and the developer could be merged in one person this is not yet the case in CSI.

\end{description}

\section{Related work}
\label{sec:relwork}
The related work is organized in two parts; first we consider the problems and guidelines in deploying MDD and Software Product Lines (SPLs), then we focus on software ecosystems.

%can be found under different prospectives: deployment of MDD approaches, in particular in large organizations, software product lines and more specifically about software ecosystems.

\subsection{Deployment of MDD and SPLs}

Baker et al. \citep{Baker2005} describe the effects of the adoption of MDD in a large company (Motorola) along 15 years. In their experience the major obstacles in adopting MDE stem to the lack of a well-defined process, lack of necessary skills and inflexibility in changing the existing culture. In CSI there were already well established processes and the company was able to form the necessary skills along the years.
%\FED{Ok ma cosa dice di interessante}
Another account from a 5-years project at the same company can be found in \citep{Foustok2007}. The author reports as difficulties the ability to cope with an increasing rate of change in the technology. In the case of CSI they were strict about dictating transitions to new technologies and they were not forced by customers to adopt particular versions or technologies. An important difference in the two experiences is the usage of UML at Motorola, while at CSI a set of DSLs was developed. 

%\FED{Ok ma come si relaziona al nostro caso?}
Fleurey et al. \citep{Fleurey2007} report about a case-study of MDE adoption. The case-study considered a time window of 10-years; the focus is on migration projects, where the benefits w.r.t. conventional techniques can be observed after an initial period, e.g. the first code could be delivered only after 10 months from project's beginning. In addition they present a cost-benefit analysis and suggest the presence of profitability threshold in terms of project size. The authors suggest that domain applications last longer than technologies, which change at a faster pace. Also in the case of CSI MDD, the key features are guaranteeing a longer life to applications and facilitating the unavoidable transitions to new technologies -- which are operated updating the generator and regenerating all the applications --.

%\FED{Ok, e allora?}
%Hen-Tov et al. \citep{HenTov2009} describe a project with enterprise software; their approach requires an initial development effort of 10 man-years.
 
While more focused on small companies, authors of this work examined the best practices for deploying MDD applications \cite{Tomassetti_Jucs_2013}. Some lessons learned for small companies apply also to this case: i) the necessity of flexibility, to achieve using DSLs and customized solution instead of framework or products off-the-shelf, ii) the importance of buying the developers committment. The first point is critical also for CSI, we suggest that software intensive companies need flexibility, to be able to tailor the development processes considering the competencies and the specificity of the company.
Organizational aspects and the importance of taking in consideration the reactions of external partners are instead peculiar to large organizations, given the impossibility for small companies to shape external factors. On one hand it means small companies have a more rigid environment to adapt to, on the other hand they have to consider less aspects in designing their MDD solution. Given that the number of users is not likely to grow up beyond the organization boundaries the MDD can be deployed with a limited investment (as it was in the first phases of the CSI case-study).

Hutchinson et al. in \cite{empAssesment_JSS} report the results of an empirical study on the assessment of MDE in industry.
Their work has two goals: identify the reasons of success or failure of MDE and understand
how MDE is actually applied in industry. They employed three forms of investigation: questionnaires, interviews, and on site observations, having as target  practitioners, MDE professionals and companies practising MDE respectively. The questionnaire has received over 250 responses from professionals (the most of them are working in Europe). 
Some of the reported findings are: (i) about two-thirds of the respondents believe that using MDE is advantageous in terms of productivity, maintability and portability (increase productivty was verified in the case of CSI, as well as portability), (ii) the majority of respondents use UML as modelling language, and a good number use in-house developed DSLs (the latter was the choice of CSI), (iii) almost three quarters of respondents think that an extra training is necessary to use MDE (we have seen that CSI invested in training), (iv) the majority of respondents agree that code generation is an important aspect of MDE productivity gain (code generation was also the choice of CSI), and (v) a little less than half of the respondents think that MDE tools are too expensive (we can confirm that CSI had to invest consistenly to create the MDD-Tools). We observed similar perceptions in a survey conducted by us \cite{Torchiano2013} except for the issue of extra-training which was not considered in our survey, however we observed that the lack of competencies is one of the problems most frequently reported by companies. Differently from the results of their survey, the cost of supporting tools is seen as a problem only by a small proportion of respondents in our sample. % while . Others are the problems hindering software modeling: too much effort required and limited usefulness.
Probably it depends on the choice to use existing tools or develop them, and on the size of the user base: we have seen that in the case of CSI a small user-base was able to use profitably an immature toolset, while it had to be considerably evolved to be adopted by a large user-base.

%\FED{The following two paragraphs have to be merged}
Hutchinson et al. \citep{Hutchinson2011} report lesson learned from adoption of MDE in three large multinational companies (a printer company, a vehicle manufacturer and a manufacturer of electronic systems). In particular, the importance of complex organizational, managerial and social factors in the success or failure of the MDE deployment.
The authors report some organizational factors that can affect the success or the failure of MDE deployment. 
The factors that can affect it positively are: (i) a progressive and iterative approach, (ii) user motivation in the MDE approach, (iii) an organizational willingness in integrating MDE in the whole organization, and (iv) having a clear business focus (where MDE is adopted as a solution for new projects). 
Instead, factors that can affect it negatively are: (i) the decision of adopting MDE being taken by IT managers, in top-down fashion and implemented ``all at once'' without developing gradually an understanding of the necessaru process changes, (ii) MDE being imposed on the developers without providing the right motivations, and (iii) an inflexible organization with a lack of integration of MDE in previous processes. 
In the CSI case all the positive factors were verified and no one of the negative ones, leading to an ideal situation.
The only common aspect with the work proposed in \cite{Hutchinson2011} concerns the motivation of developers. %The corresponding finding lies in the problems reported in Figure \ref{fig:problems} (refusal from developers and refusal from management).

Mohagheghi et al.~\cite{Mohagheghi2012_JSS} interviewed -- using convenience sampling -- developers from four companies involved in an initiative called MODELPLEX. They examined the factors affecting adoption of MDE. Regarding usefulness they found uncertain results: most participants recognize the usefulness of models but they are not sure about the impact on the quality of the final product or the effects on productivity. MDE is perceived as not simple: its complexity makes it viable for engineers but not for non technical people. This finding is confirmed by our results reported in~\cite{torchianoESEMsurv2011_JSS,tomassetti2012_JSS}. They show that only in a few cases business experts are involved during modelling tasks. Also in our case-study the introduction of MDD led to re-consider the different roles and their involvement in the development phases (see Sect. \ref{sec:discussion} about Roles). 

Regarding compatibility with the existing development process the companies complained about the lack of standards and the consequent lock-in effect. All interviewed companies reported some problems in integrating their existing approaches with MDE. Tools could have been part of their problems, them being not considered satisfying by a part of the sample. In particular, some participants expressed several concerns about the scalability of the MDE approach to large projects (this could be related to the motifs \emph{Integration} and \emph{Support}). Advantages reported are limited to the usefulness for documentation and communication purposes. Major reasons preventing adoption of MDE are the immaturity of tools and processes  as well as the lack of competencies. %Such latter conclusions are largely consistent with our findings.
%his study contains some threats to validity: companies involved in this study performed mainly exploratory applications of modelling and the goal of their usage is not clear: was it just for documentation purposes or did they generated some code? The number of companies involved is also very low. In respect to this work we tried to capture a more complete picture of the adoption of modeling for different purposes and at different levels of maturity.

%\cite{Trask2011}
Catal \cite{Catal2009} discusses some of the barriers to the adoption of Software Product Line Engineering (SPLE). Among the other findindgs, the author suggest to consider SPLE as mean to obtain resusability at different levels: not only of implementation artifacts, but also of documentation, tests, practictes and other complimentary elements. According to Catal some of the problems derive from an unclear terminilogy (with duplicate terms used in US and Europe) and a lack of resources to learn how to implement SPLEs. The necessity of deep changes in the organization's process make seniors to refrain and resist the migration.  

Authors of this work partecipated on an industrial survey about MDD adoption in Italian companies \cite{Torchiano2013}. Among other findings, from our survey emerges that the most common problems in deploying MDD are the size of the effort required, the necessity to prove the usefulness of the solution designed, the lack of competencies and proper tools, the missing support from management. In this case the company solved this issues through a slow evolution which made possible to limit the initial investment and to progressively buy-in management and developers support. The size of the company made possible to face the investments necessary both to design the solution but also to grow the necessary competencies. We think that this partial transfer of competencies to external companies and the availability of the toolset to other companies could help other companies part of the district to adopt MDD, lowering the entry-barriers.

\subsection{Software ecosystems}

%\cite{Herold2008}
%\cite{Lentz2007}

%\FED{Ok ma a cosa serve?}
In 2009 Jansen et al. \cite{Jansen2009} proposed a research agenda for software ecosystems (SECOs). The three prospectives considered are the \emph{software ecosystem level}, the \emph{software supply network level} and the \emph{software vendor level}. On the \emph{software ecosystem level} strategies are implemented to keep the ecosystem profitable for all the participating actors; our motif \emph{RoI for adopters} clearly operates at this level, also all the motifs related to efficiency could be considered as related to this level because the efficiency of the tools positevely affect all the participants in the ecosystem. On the \emph{software supply network level} relations with suppliers and buyers are considered: orchestration between partners at different part in the process-chain is administered at this level; the motif \emph{automatic enforcement} permits to reduce the cost of verifying the conformity of the products with the requirements, an important aspect in managing the relations with suppliers. Finally at the \emph{software vendor level} actors considers the effect of the SECO on their own products catalog.
For each prospectives different challenges are reported. Our work addresses a challenge at  the \emph{software ecosystem level}, in particular \emph{Developing policies and strategies
within SECOs for SECO orchestration}.

Barbosa and Alves \cite{Barbosa2011} presented a systemic mapping study on SECOs. Authors included 44 relevant papers. The study, being conducted in 2011, include papers up to year 2010. The number of papers show an increase of interest in SECO in the most recent years considered (2010-2011). Of these 44 papers 4 are focusing on SECO and SPLs; it seems to suggest that more research is needed in this direction, and this case apply to MDD-Tools.

Angeren et al. \cite{vanAngeren2011} performed a survey in the Dutch software industry about the ecosystems companies are working in. Data was collected from bachelor students according to given schemas. In the end the data of 17 companies was considered. Authors individuate four different categories of components, obtained from the combination of two dimensions (critical/non critical, core/contextual). For each category they report which factors influence the relations with suppliers. Factors are: level of intimacy, continuity, visibility within the marked, niche creation, product \& license type, support \& maintenance. According to their schema the MDD-Tools could be considered as a critical core component, therefore all the factors would be relevant.

%Van den Berk et al.\cite{vandenBerk2010} proposes the SECO Strategy Assessment Model (SECO-SAM) to assess decisions taken rabout the evolution of software ecosystems and forecast possible outcome of new strategies. Authors states that the organization responsible for administrating the platform (called the hub) is the element which has most influence

%\FED{How it is related?}
%McGregor proposes a method for analyzing software product line ecosystems \cite{McGregor2010}. In particular the method focuses in on considering the flow of information between the different actors (both intra-organization and extra organization actors) differentiation between transfers (exchanges without any cost or effort) and transactions. 

Bosch \cite{Bosch2009} examines how successful software product lines can evolve to large ecosystems. He considers in particular the case in which a company starts using a SPL for developing its own products and later open it to other actors. The SPL evolve into a software ecosystem when the platform starts to be used outside the organization boundaries. According to Bosch the two main reasons motivating the transactions are: i) the impossibility for the company to sustain alone the R\&D costs, ii) the mass customization necessary in certain sectors (e.g., web) for application of the SPL to different customers. In the case of MDD-Tools the major advantages guiding the transitions were the need for standardization of development times, cost and platforms, together with an easier governance of outsourced components. Bosch proposes also a taxonomy of software ecosystems, considering the category (end-user programming, application, OS) and the platform (desktop, web, mobile). We would suggest to add the "tools based ecosystem" which is somewhat similar to the end-user programming ecosystem presented by Bosch (in both case the central element is the tool used to develop applications) but some aspects are fundamentally different: in particular "end-user programming" ecosystems are based on the fact that few investments are needed for the adoption while out suggested category (tools-based) would imply relevant cost for teaching and supporting the developers adopting the tool. In both cases the deep customization of the applications is based on DSLs. 

%\cite{vanAngeren2011b}

Hannsen \cite{Hanssen2012}, similarly to us, studied the evolution of a Software-Product-Line centered ecosystem for a period of five years. While in their case-study the ecosystem is born in the period of interest and the SPL was adopted, in our case-study both the SPL and the ecosystem were there from the beginning; however in the period of interest there was a critical technological change in the implementation of the SPL. The company considered by Hansen is smaller than CSI (260 people employed against 1200) but it is a worlwide distributed company, instead of a locally based company as CSI. The author describes in details the involvement of the customers as an important aspect of the ecosystem. This strong attention to stakeholders seem to derive from the transition to the Evo method \cite{Gilb2005}. Engaging customers became part of the culture of the company, with reflections at different levels. While initially it required an active effort to involve stakeholders, later the company was able to trigger a strong interest and constitute a pool of very active partners. The company was also able to act as a catalyst for the 60 external organizations which are basing their business of the product line. They did it organizing conferences, opening a web portal and nurtuting the network of partners; some of these ideas could be applied by CSI in the future to strengthen the network of MDD-Tools' users. This is particularly good for those external organizations because they can easily get in touch with a large pool of established users. An efficient integration technology was considered an enabled for this set of extensions, customizations and interoperable solutions.
The author presents also a set of theoretical propositions defining a software ecosystem. Among them we note i) the presence of a central referent organization, acting as hub of the software ecosystem. In our case CSI is clearly playing this role; ii) the enabling role of a particular technology, in our case study MDD; iii) the shift to a shared responsibility model on the development and control of the ecosystem, this element is present also in our case-study with the release of MDD-Tools as an open-source platform.

Manikas and Hansen \cite{Manikas2013} conducted a systematic review of the literature about software ecosystems (SECOs). Authors based their method on the guidelines from Kitchenham and Charters \cite{Kitchenham2007}. They considered 420 papers and included 90.  Here we report a summary of their findings: i) there are different definitions of software ecosystem being used, the ones most widely referred come from Jansen et al. \cite{Jansen2009} or Bosch et. al \cite{Bosch2009, Bosch2010a,Bosch2010b}; ii) the number of papers published on the topic is increasing significantly, raising from 3 papers published in 2007 and 2008 to the 32 published in 2010 and 2011 (the papers' extraction was performed during June 2012); iii) almost half of the papers are reports, while very few use empirical methods. This is an aspect common to many other fields of the SE; iv) papers have a sort of equally distributed focus between SE, business and management, and ecosystems relationships; v) half of the papers refer to an existing SECO.

Bosch and Bosch-Sijtsema discuss about the impact of software product lines and ecosystems \cite{Bosch2010a}. They suggest that large-scale software development is hindered by a too much integration-centric approach while switching to a composition-oriented approach would significantly simplify it. To reach this goal the factors motivating strong and close interconnections are examined in this work. Initially authors present three trends affecting large software development (SPLs, global development and SECOs). Based on their experience and in particular on three case study companies authors present problems common to software intensive companies. Some of the problems derive from the software architectures, making costly the integration and difficult the indipendent evolution of parts of the system. Other problems could derive by engineering practices and from the R\&D organization.
Finally authors present five approaches to facilitate the transition to a more composition-oriented system: i) consider integration during development, ii) release groupings, iii) release traits, iv) independent deployment, and v) open ecosystem.

Lungu et al. \cite{Lungu2010} presented a tool to visualize the evolution of SECOs. The tools is named Small Project Observatory and it is an online visualization tool focusing on super-repositories (federation of repositories). It could be used as a supporting tool in studying the patterns of evolutions of applications and improve reusability through the definition of processes common to the ecosystem. That would help to pose the basis for a transition to more structured approaches, as the one proposed by CSI.

Kilamo et al. \cite{Kilamo2012} list a set of guidelines for successfully release as open-source a proprietary software system and grow a proper ecosystem. Authors derived those guidelines from applying a previously depicted process (the OSCOMM process framework) to four different case-studies. The OSCOMM framework consists of three phases: i) an evaluation of the readiness of the project for being opened, ii) open source engineering the product, and, iii) measuring the ecosystem once the project is open.

%\cite{Bosch2012}

%\cite{Kettunen2005}

\section{Conclusions}
\label{sec:conclusions}
In this paper we summarized the five years long evolution of a software ecosystem centered on a large IT organization (CSI-Piemonte). In particular we presented eight motifs that played a significant role in the successful deployment of a paradigm change in such a complex ecosystem. 

Although the motifs are derived from one single study -- which is a limitation to their validity --, this particular study spanned across many years and considered hundreds of projects, we therefore think that the distilled motifs represent a valuable contribution for deriving best practices in performing paradigm shifting within a software ecosystem.

Some of the motifs presented played a role in creating an \emph{efficient} ecosystem, while other were important to favor the \emph{diffusion} of the paradigm change. Some of them played a role in both aspects. %In Table \ref{tab:motifEffects} we reported a list of the motifs, the era in which they were put in action and the aspects they affected.
The diffusion was initially favoured by \emph{Incremental adoption} on voluntary basis, then offering the necessary \emph{Support} and \emph{Integration} in the toolchain. In the later eras complementary aspects had to be considered like a proper \emph{RoI for adopters} and a shared responsibility for the solution, obtained through \emph{Distributed development}.
The diffusion was also indirectly favored by realizing an efficient ecosystem first of all with the fundamental choice of performing \emph{Toolsmithing} (with all the implications at an ecosystem level). \emph{Integration} could be also considered an element which brought to an efficient solution. Finally investments in \emph{Automatic enforcement} and \emph{Generated code quality} were particularly rewarding (and sustainable) thanks to the economies of scale of the adoption at an ecosystem level.

With this interpretative work we attempt to propose key motifs for a paradigm shit in a software ecosystem. We believe more interpretative works are needed in the area, to find similarities and variabilities with analogous evolutions in other ecosystems. At that stage generalization would be possible by analyzing the different single experiences.

In particular we believe it would be useful to compare this work with other regional software development ecosystem studying the impact of introducing a rarely used paradigm to a large number of developers in a specific area. The effects in terms of know-how diffusion, the number of initiatives which independently spread out from this effort have yet to be evaluated. We believe that local ecosystems can lead to an unexpected number of interactions which have to be fully studied and understood.

\section*{Acknowledgments}
The author would like to thank all the persons who offered their feedback. In particular Claudio Parodi and Marco Boz offered important insights which helped to complete and clarify our report.



%% References


\bibliographystyle{model1-num-names}
\bibliography{biblio}






% \end{thebibliography}

\end{document}